\documentclass[iop,apj]{emulateapj}
\usepackage{graphicx,url}

\shorttitle{Spectroscopy of VV124}
\shortauthors{Kirby, Cohen, \& Bellazzini}
\slugcomment{Accepted to ApJ on 2012 March 19}

\newcommand{\vsyserr}{2.21}
\newcommand{\meanv}{-29.1}
\newcommand{\meanverr}{1.3}

\newcommand{\medianvrerr}{4.1}
\newcommand{\sigmav}{ 9.4}
\newcommand{\sigmaverr}{1.0}
\newcommand{\mass}{2.1}
\newcommand{\masserr}{0.2}
\newcommand{\ml}{5.2}
\newcommand{\mlratio}{1.5}
\newcommand{\mlerr}{1.1}
\newcommand{\mratio}{4.5}
\newcommand{\mratioerr}{1.9}

\newcommand{\vskewness}{ 0.14}
\newcommand{\vskewnesserr}{0.29}
\newcommand{\vkurtosis}{-0.26}
\newcommand{\vkurtosiserr}{0.58}
\newcommand{\fehmin}{-2.9}
\newcommand{\fehmax}{-0.7}
\newcommand{\fehmean}{-1.58}
\newcommand{\fehmeanerr}{0.06}

\newcommand{\medianfeherr}{0.23}
\newcommand{\fehslope}{-0.22}
\newcommand{\fehslopeerr}{0.05}
\newcommand{\probratio}{4.1}

\newcommand{\nmember}{67}
\newcommand{\nbad}{13}
\newcommand{\nvexclude}{6}

\newcommand{\nvdiff}{57}

\newcommand{\avgvdiff}{3.5}
\newcommand{\avgvdifferr}{8.6}

\begin{document}

\title{The Dynamics and Metallicity Distribution of the Distant Dwarf Galaxy VV124\altaffilmark{*}}

\author{Evan~N.~Kirby\altaffilmark{1,2},
  Judith~G.~Cohen\altaffilmark{1}, and
  Michele~Bellazzini\altaffilmark{3}}

\altaffiltext{*}{The data presented herein were obtained at the
  W.~M.~Keck Observatory, which is operated as a scientific
  partnership among the California Institute of Technology, the
  University of California and the National Aeronautics and Space
  Administration. The Observatory was made possible by the generous
  financial support of the W.~M.~Keck Foundation.}
\altaffiltext{1}{California Institute of Technology, 1200
  E.\ California Blvd., MC 249-17, Pasadena, CA 91125, USA}
\altaffiltext{2}{Hubble Fellow.}  \altaffiltext{3}{INAF - Osservatorio
  Astronomico di Bologna, via Ranzani 1, 40127 Bologna, Italy}


%%%%%%%%%%%%%%%%%%%%%%%%%%%%%%%%%
%%%%%%%%%    ABSTRACT    %%%%%%%%
%%%%%%%%%%%%%%%%%%%%%%%%%%%%%%%%%

\begin{abstract}

VV124 (UGC~4879) is an isolated, dwarf irregular/dwarf spheroidal
(dIrr/dSph) transition-type galaxy at a distance of 1.36~Mpc.
Previous low-resolution spectroscopy yielded inconsistent radial
velocities for different components of the galaxy, and photometry
hinted at the presence of a stellar disk.  In order to quantify the
stellar dynamics, we observed individual red giants in VV124 with the
Keck/DEIMOS spectrograph.  We validated members based on their
positions in the color-magnitude diagram, radial velocities, and
spectral features.  Our sample contains \nmember\ members.  The
average radial velocity is $\langle v_r \rangle = \meanv \pm
\meanverr$~km~s$^{-1}$, in agreement with the previous radio
measurements of \ion{H}{1} gas.  The velocity distribution is
Gaussian, indicating that VV124 is supported primarily by velocity
dispersion inside a radius of 1.5~kpc.  Outside that radius, our
measurements provide only an upper limit of $\avgvdifferr$~km~s$^{-1}$
on any rotation in the photometric disk-like feature.  The velocity
dispersion is $\sigma_v = \sigmav \pm \sigmaverr$~km~s$^{-1}$, from
which we inferred a mass of $M_{1/2} = (\mass \pm \masserr) \times
10^7~M_{\sun}$ and a mass-to-light ratio of $(M/L_V)_{1/2} = \ml \pm
\mlerr~M_{\sun}/L_{\sun}$, both measured within the half-light radius.
Thus, VV124 contains dark matter.  We also measured the metallicity
distribution from neutral iron lines.  The average metallicity,
$\langle {\rm [Fe/H]} \rangle = \fehmean \pm \fehmeanerr$, is
consistent with the mass-metallicity relation defined by dwarf
spheroidal galaxies.  The dynamics and metallicity distribution of
VV124 appear similar to dSphs of similar stellar mass.

\end{abstract}

%%%%%%%%%%%%%%%%%%%%%%%%%%%%%%%%%
%%%%%%%%%   SECTION 1   %%%%%%%%%
%%%%%%%%%%%%%%%%%%%%%%%%%%%%%%%%%

\section{Introduction}
\label{sec:intro}

The dwarf galaxy VV124 is one of the ``newest'' members of the local
universe.  The discovery of VV124 was first published in the
\citet{vv59} Atlas and Catalog of Interacting Galaxies, though the
galaxy does not have an interacting neighbor.  The CfA galaxy redshift
survey \citep{huc83} reported its redshift as $cz = 600 \pm
100$~km~s$^{-1}$.  In the Hubble flow, this redshift corresponds to a
distance of $\sim 10$~Mpc.  At that distance, VV124 would have
belonged to a group of three galaxies \citep{tur76,gel83}.  However,
\citet{kop08} noticed resolved stars in Sloan Digital Sky Survey
images \citep{ade07}.  \citeauthor{kop08}\ imaged VV124 to a depth of
$V = 25.6$, and they confirmed that it is in fact much nearer than
10~Mpc.  They measured a distance of $1.1 \pm 0.1$~Mpc based on the
tip of the red giant branch (TRGB)\@.  This new distance placed the
galaxy on the periphery of the Local Group.  Despite its proximity to
the group, VV124 is extremely isolated.  Its nearest neighbor, Leo~A,
is another isolated dwarf galaxy 700~kpc away \citep{jac11}.

%The Uppsala Galaxy Catalog \citep{nil73} renamed the galaxy UGC~4879,
%and it listed it as an irregular galaxy.  The Third Reference
%Catalogue of Bright Galaxies \citep{dev91} refined the classification
%to IAm: an irregular, non-barred, Magellanic-type galaxy.  Its
%luminosity class is V--VI (low-surface brightness with some
%structure).

The revision to VV124's distance drastically affected its inferred
star formation history.  From low-resolution spectroscopy,
\citet{jam04} measured its H$\alpha$ star formation rate as $(5.0 \pm
2.4) \times 10^{-3}~M_{\sun}~\rm{yr}^{-1}$, but that rate was based on
a distance of 10.5~Mpc.  Corrected to the true distance of 1.36~Mpc
\citep{jac11}, the rate dropped to $8.4 \times
10^{-5}~M_{\sun}~\rm{yr}^{-1}$ \citep[a point also noted by][hereafter
  B11a]{bel11a}.  The new distance also inspired deep photometry in
addition to that obtained by \citet{kop08}.
\citeauthor*{bel11a}\ obtained wide-field $g,r$ photometry of VV124 at
the Large Binocular Telescope (LBT) to study the structure of the
galaxy in detail.  \citet{jac11} obtained even deeper $V$ and $I$
photometry over a smaller field with the {\it Hubble Space Telescope}
(HST) Advanced Camera for Surveys.  They determined that $93\%$ of the
stars are older than 10~Gyr.  The rest are younger than 1~Gyr.
\citet{bel11b} used the same HST images as \citeauthor{jac11}\ to show
that the oldest, most metal-poor population is significantly more
extended than the younger stars.  The galaxy-averaged star formation
rates that \citeauthor{jac11}\ derived were $(3.0 \pm 0.2) \times
10^{-4}~M_{\sun}~\rm{yr}^{-1}$ over the last 0.5~Gyr and $(8.0 \pm
0.5) \times 10^{-4}~M_{\sun}~\rm{yr}^{-1}$ over the last 1~Gyr.
Together, the HST photometry and H$\alpha$ spectroscopy indicated that
VV124 is a typically old dwarf galaxy that experienced a small burst
of star formation $\sim 1$~Gyr ago.  %The star formation rate has
%slowly declined since then.

\citet{tik10} expanded \citeauthor{kop08}'s (\citeyear{kop08}) study.
They classified VV124 as transition-type galaxy, caught between being
an dIrr and a dwarf spheroidal galaxy (dSph).  They noticed that the
young, blue stars lie in a thin, disk-like structure, whereas the
older, red giants occupy a thicker configuration.  However, the small
number of very young stars was inconsistent with an irregular-only
classification.  Therefore, they concluded that VV124 is a rare
dIrr/dSph galaxy.  \citet{gre03} listed only five other such members
of the Local Group: Antlia, Aquarius, Phoenix, Pisces, and KKR~25.
This class of galaxies is especially interesting because their
existence supports the theory that dSphs are dIrrs that lost their gas
through ram pressure stripping and tidal interactions with other
galaxies \citep{may01a,may01b,may06,may07}.  There is strong
observational support for the removal of gas from dwarf galaxies in
dense environments \citep[e.g.,][]{geh06,grc09}, and the similarity of
the star formation histories except in the past Gyr of dIrrs, dSphs,
and dIrr/dSphs further supports the transformation of dIrrs into dSphs
\citep{wei11}.  Although there are counterexamples, such as the
comparatively young Leo~A dIrr \citep{col07}, the star formation
history of VV124 less recent than 1~Gyr appears much like other dIrrs
and dSphs.  Except for the recent burst, \citet{jac11} reported that
almost all of its stars are ancient, like nearly all dSphs.

The dynamics of VV124 have been unclear.  Despite the corrected
redshift, measurements of the galaxy's radial velocity have been
inconsistent.  \citet{kop08} and \citet{tik10} obtained low-resolution
spectroscopy of individual stars, resolved \ion{H}{2} regions, diffuse
stellar light, and diffuse emission.  The velocities varied widely,
but they seemed to cluster around $v_r = -70$~km~s$^{-1}$.
\citeauthor*{bel11a}\ measured the \ion{H}{1} velocity structure with
the Westerbork Synthesis Radio Telescope.  They measured a
heliocentric radial velocity of $v_r = -25 \pm 4$~km~s$^{-1}$ with a
dispersion of $\sim 11$~km~s$^{-1}$ and a full range of about $-45$ to
0~km~s$^{-1}$.  \citeauthor*{bel11a}\ also measured the velocities of
a blue supergiant and a young star cluster, but the cluster velocity
was inconsistent with (more negative than) the \ion{H}{1} velocity by
$2.7\sigma$.  So far, the discrepancy between the velocities of the
neutral gas and younger populations has not been resolved.

\citeauthor*{bel11a}\ also observed what appears to be a stellar disk.
VV124 is elongated ($\epsilon = 0.44 \pm 0.04$, ${\rm PA} = 84\arcdeg
\pm 11\arcdeg$), but its shape cannot be described by a single
ellipsoid.  There is an even more elongated component whose major axis
is aligned with the main spheroid.  The second component, which
\citeauthor*{bel11a}\ called ``wings,'' resembles a stellar disk,
which should rotate coherently, unlike the spheroid.  As an
alternative, \citeauthor*{bel11a}\ also proposed that the wings might
be tidal tails instead of a disk.  However, VV124's extreme isolation
makes strong tidal interaction unlikely.

In order to provide closure to the discrepancies in velocity
measurements and to test for rotation, we observed red giants in VV124
with the Deep Extragalactic Imaging Multi-Object Spectrograph
\citep[DEIMOS,][]{fab03} on the Keck~II telescope.  The median
velocity uncertainty of the individual stellar measurements is
\medianvrerr~km~s$^{-1}$, which is small enough to test for the
presence of dark matter.  The spectra are also of sufficient quality
to measure metallicities from neutral iron lines.  The median
uncertainty in metallicity is \medianfeherr~dex.  This article
presents the results of our dynamics and metallicity studies.

%%%%%%%%%%%%%%%%%%%%%%%%%%%%%%%%%
%%%%%%%%%   SECTION 2   %%%%%%%%%
%%%%%%%%%%%%%%%%%%%%%%%%%%%%%%%%%

\section{Observations and Reductions}
\label{sec:obs}

\subsection{Target selection}
\label{sec:selection}

Targets were selected from \citeauthor*{bel11a}'s photometric catalog.
DEIMOS slitlets cannot overlap in the dispersion direction, and the
VV124 field is dense enough that we needed to prioritize the targets
for placement on the DEIMOS slitmasks.  The targets were prioritized
by magnitude and physical location with respect to the center of
VV124\@.  Brighter targets were preferred for higher signal-to-noise
ratios (SNRs).  We defined $X$ as the distance along the major axis
and $Y$ as distance along the minor axis.  This is the same
nomenclature introduced by \citeauthor*{bel11a}.

We used the
\texttt{dsimulator}\footnote{\url{http://www.ucolick.org/$\sim$phillips/deimos\_ref/masks.html}}
program to select targets for placement on DEIMOS slitmasks.  The
program maximizes the number of high-priority targets within
user-defined constraints.  We required slitlets to be at least 4''
long and separated by 0.35'' for adequate sky subtraction.  Slit
widths were 0.7'', a choice that balanced observing efficiency with
spectral resolution, which corresponds to precision in the radial
velocity and metallicity measurements.

We prioritized targets for spectroscopy in the following order:

\begin{enumerate}
\item Brighter red giant branch (RGB) candidates near the major axis ($V \le 23.7$ and $|Y| \le 2'$)
\item Fainter RGB candidates near the major axis ($V > 23.7$ and $|Y| \le 2'$)
\item Bright, blue non-RGB stars near the major axis ($V \le 23.7$, $g-r < 1.1$, and $|Y| \le 2'$)
\item RGB candidates away from the major axis ($|Y| > 2'$)
\item Any other non-RGB candidates
\end{enumerate}

\noindent RGB candidates are defined as $r > 22$ and $0.7 < g-r <
1.4$.

One pair of slitmasks comprised our observations.  The slitmasks,
subtending roughly $16' \times 4'$, were roughly centered on the
galaxy at position angles of $86\arcdeg$ and $-94\arcdeg$ (i.e.,
rotated $180\arcdeg$).  The target lists for the two slitmasks were
almost identical.  This ensured that we had two independent
measurements of the radial velocities for almost all of the stars.
The target list included 131 unique targets, of which 106 were
included on both slitmasks.

\begin{figure}[t!]
\centering
\includegraphics[width=\columnwidth]{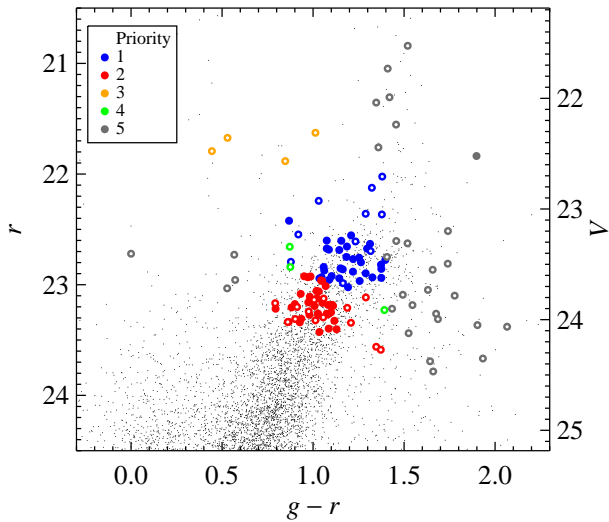}
\caption{CMD from LBT photometry \protect (\citeauthor*{bel11a}).
  Colored points identify stars selected for DEIMOS spectroscopy.  The
  color signifies the priority of the star for spectroscopic selection
  (Section~\ref{sec:selection}).  Filled points are spectroscopically
  confirmed member stars, and hollow points are non-members.  The
  right axis gives the approximate Cousins $V$
  magnitude.\label{fig:cmd}}
\end{figure}

\begin{figure}[t!]
\centering
\includegraphics[width=\columnwidth]{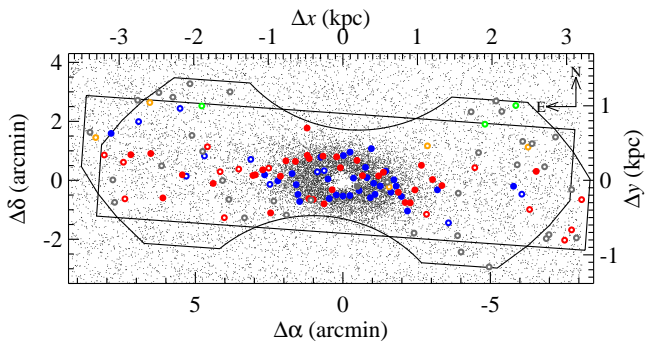}
\caption{DEIMOS slitmask footprints overlaid on the LBT astrometric
  catalog \protect (\citeauthor*{bel11a}).  The colors and types of
  points have the same meanings as in Figure~\ref{fig:cmd}.  The axis
  labels give the displacement from the center of VV124 ($\alpha_0 =
  9^{\rm h}16^{\rm m}03^{\rm s}$, $\delta_0 = +52\arcdeg 50\arcmin
  31\arcsec$).  The top and right axes give this displacement in kpc
  for an assumed distance of 1.36~Mpc \protect \citep{jac11}.  The
  positions and shapes of the two DEIMOS slitmasks are
  outlined.\label{fig:coords}}
\end{figure}

Figure~\ref{fig:cmd} shows the color-magnitude diagram (CMD) for
VV124\@.  Spectroscopic targets are identified by colored points.
Filled points indicate spectroscopically confirmed member stars (see
Section~\ref{sec:membership}).  The colors correspond to the
priorities in the list above.  Figure~\ref{fig:coords} shows the
coordinates of the stars with the same color coding.

\subsection{Observations}

\begin{deluxetable*}{lccccc}
\tablewidth{0pt}
\tablecolumns{6}
\tablecaption{DEIMOS Observations\label{tab:obslog}}
\tablehead{\colhead{Mask} & \colhead{Seeing ('')} & \colhead{Targets} & \colhead{Members} & \colhead{Total Exp. Time (s)} & \colhead{Individual Exp. Times (s)}}
\startdata
vv124a & 0.57--0.87 & 121 & 61 & 13200 & $6 \times 1800$, $2 \times 1200$ \\
vv124b & 0.55--0.65 & 120 & 64 & 13600 & $6 \times 1800$, 1600, 1200 \\
\enddata
\end{deluxetable*}

We observed the two slitmasks, called vv124a and vv124b, on 2011
January 30 (UT).  Table~\ref{tab:obslog} lists the exposure times,
seeing, number of targets, and number of VV124 members for each
slitmask.  The following night was also allocated to this project, but
high humidity and fog on that night prevented any observing of
VV124\@.

We used the 1200~lines~mm$^{-1}$ grating centered at 7800~\AA\@.  This
grating, combined with the slit widths of 0.7'', yielded a line spread
profile with a FWHM of 1.2~\AA\@.  The resolving power is about $R
\sim 7000$ near the \ion{Ca}{2} triplet (CaT) at 8500~\AA\@.  We
obtained a 1~sec exposure of internal Ne, Ar, Kr, and Xe arc lamps for
wavelength calibration and three 12~sec exposures of an internal
quartz lamp for flat fielding.  Although DEIMOS is mounted on the
Nasmyth platform of Keck~II, it rotates to track the sky.  An active
flexure compensation system maintains stability to within 0.2~pixels,
corresponding to 3~$\mu$m on the focal plane, which is equivalent to
0.02'' in the spatial direction and to 0.07~\AA\ in the dispersion
direction.

Individual exposure times for the VV124 slitmasks ranged from 20 to
30~minutes.  Most of the exposures were 30~minutes.  We checked the
alignment of the slitmask about once per hour by placing the grating
in zeroth order.  In this direct imaging mode, bright alignment stars
aligned with 4'' boxes in the slitmask.  We measured the seeing by
fitting Gaussian profiles to the point spread functions of the
alignment stars.  Seeing remained low and stable for the entire night
(see Table~\ref{tab:obslog}).

\subsection{Reductions}

We reduced the raw images into one-dimensional spectra with the
\texttt{spec2d}\footnote{\url{http://deep.berkeley.edu/spec2d/}}
software developed by the Deep Extragalactic Evolutionary Probe~2 team
\citep[DEEP2,][]{dav03}.  The software excised the footprint of each
spectrally dispersed slitlet from each exposure of the flat field, arc
lamps, and science targets.  Because the DEIMOS focal plane is curved,
the program rectified the slitlet images by tracing and straightening
both the edges of the flat field and the roughly orthogonal arc lines.
Then, the program generated a flat field image from the three
exposures of the quartz lamp.  This flat field was used to remove the
slit response function, dust, and cosmetic imperfections in the
science images.  The \texttt{spec2d} software also fitted a
fifth-order Legendre polynomial in wavelength to the pixel positions
of arc lines.  For each exposure of each slitlet, \texttt{spec2d}
computed the sky emission as a function of wavelength and subtracted
it from the two-dimensional spectrum.  All of the science exposures
for each slitmask were averaged with inverse variance weighting and
cosmic ray rejection.  Finally, the object spectrum was extracted with
optimal weighting \citep{hor86}.  The final result was a clean,
one-dimensional spectrum, wavelength-calibrated to a precision of
0.01~\AA\@.  The spectrum was not flux calibrated, and it did not need
to be for our purposes.

Some spectra suffered from artifacts or such low SNR that measurement
of the radial velocity was not possible.  There were \nbad\ such
spectra.  They are included and identified in Table~\ref{tab:catalog},
but we do not consider them further.

The \texttt{spec2d} software also computed the error spectrum from
Poisson statistics of the raw frame.  The errors were adjusted
appropriately during flat fielding, and the one-dimensional error
spectrum was extracted identically to the flux spectrum.  We used the
error spectra for estimation of uncertainties in radial velocities
(Section~\ref{sec:rv}) and metallicities
(Section~\ref{sec:metallicity}).

As mentioned in Section~\ref{sec:selection}, the two slitmasks
included 106 duplicate targets.  We measured the precision of our
velocity measurements using the independently measured radial
velocities from the two slitmasks.  However, for the final list of
radial velocities, we used a combined spectrum.  We summed the two
one-dimensional spectra for each duplicate object with inverse
variance weighting.

We removed telluric absorption using archived DEIMOS spectra of hot
stars.  We used the same hot star spectra and spectral division
procedure as \citet{kir08}.  However, our science exposures took place
over an entire night.  The varying airmass and water vapor made it
impossible to remove the telluric absorption accurately.  Regions of
moderately strong telluric absorption were not considered for
metallicity measurements.

The metallicity measurements discussed in
Section~\ref{sec:metallicity} required normalization of the spectral
continuum.  We followed the continuum normalization procedure
described by \citet[][their Section~3.4]{kir09}.  After normalization,
we calculated the SNR of each spectrum.  First, we computed the median
absolute deviation (m.a.d.)\ from the continuum of pixels in
``continuum regions'' \citep[defined by][]{kir08} in the rest
wavelength ranges 7400--7500~\AA, 7750--8100~\AA, and
8400--8900~\AA\@.  These selections excluded most telluric absorption
and TiO bands.  Then, we removed all pixels that exceeded five times
the m.a.d.\ and re-calculated the m.a.d.\ with this slightly trimmed
set of pixels.  The result was the SNR per pixel.  To convert to SNR
per \AA, we multiplied by $\sqrt{1 / 0.33}$, where 0.33 is the pixel
scale in \AA\ per pixel.  This measurement of SNR is sensitive to
inaccuracies in the continuum placement, which will artificially
decrease the SNR\@.  However, errors in continuum placement dominate
the SNR calculation only at $\rm{SNR} \ga 60~\rm{\AA}^{-1}$, which is
much higher than any member star in our sample.

%%%%%%%%%%%%%%%%%%%%%%%%%%%%%%%%%
%%%%%%%%%   SECTION 3   %%%%%%%%%
%%%%%%%%%%%%%%%%%%%%%%%%%%%%%%%%%

\section{Radial Velocity Measurements}
\label{sec:rv}

\begin{figure}[t!]
\centering
\includegraphics[width=\columnwidth]{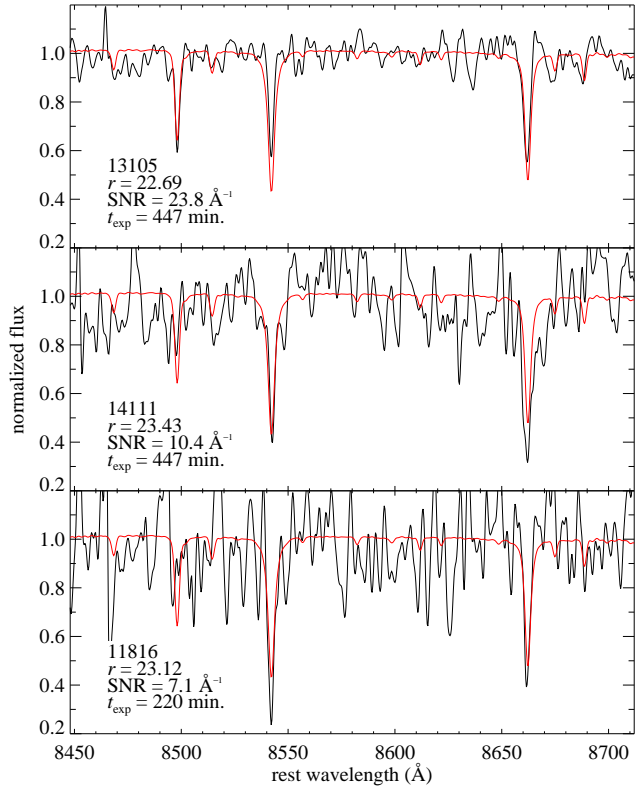}
\caption{DEIMOS spectra of three red giant members of VV124\@.  Only a
  small portion of the spectrum around the CaT is shown.  For display,
  the spectra have been normalized to the continuum and smoothed with
  a Gaussian kernel ($\sigma = 0.7$~\AA).  The radial velocity
  template spectrum, provided by \protect \citet{sim07}, is shown in
  red in all three panels.  The top and bottom panels show the spectra
  with the highest and lowest SNRs in our sample.  The middle panel
  shows the faintest member star in our sample.\label{fig:example}}
\end{figure}

We measured radial velocities and their uncertainties following the
procedure of \citet{sim07}.  First, we measured the radial velocity of
the star by cross-correlating the observed spectrum with a library of
template stars and galaxies observed with DEIMOS\@.  The template
spectra were the same as those used by \citeauthor{sim07}.  They were
observed with a nearly identical configuration of DEIMOS and therefore
have the same resolution as our science spectra.  The strong
atmospheric A and B~bands (7588--7700~\AA\ and 6862--6950~\AA) were
excluded from the cross-correlation.  Three weaker telluric bands
(7167--7315~\AA, 8210--8325~\AA, and 8900--9200~\AA) were given
weights of 10\% of other pixels in the cross-correlation.  We adopted
the velocity corresponding to the cross-correlation peak of the
template with the lowest $\chi^2$ when compared to the observed
spectrum.  We call this velocity $v_{\rm obj}$.
Figure~\ref{fig:example} shows typical spectra of VV124 member stars
along with the best-fitting template spectra.

Next, we computed a correction for mis-centering of the star in the
slitlet \citep{soh07}.  For this step, we used the observed spectrum
before correction for telluric absorption.  We cross-correlated this
spectrum with the spectrum of HR~1641, a hot, rapidly rotating star.
Only regions subject to significant telluric absorption \citep[the
  same regions used by][]{sim07} were considered in the
cross-correlation.  We call the velocity obtained from this
cross-correlation $v_{\rm tell}$.  The radial velocity of the star is
$v_r = v_{\rm obj} - v_{\rm tell} - v_{\rm helio}$, where $v_{\rm
  helio}$ is the heliocentric velocity correction appropriate for Keck
Observatory on the observation date.

We estimated uncertainties in $v_r$ by Monte Carlo resampling of the
spectra.  For each of 1000 realizations, we added Gaussian random
noise to the spectrum.  The dispersion of the Gaussian probability
distribution at each pixel was equal to the estimated Poisson error of
that pixel.  Then, we recomputed $v_{\rm obj}$ and $v_{\rm tell}$ for
each realization.  For computational efficiency, we considered only
the template spectrum that best matched the original spectrum.  The
standard deviation of the 1000 measurements of $v_r$ is $\sigma_{\rm
  MC}$.

\begin{figure}[t!]
\centering
\includegraphics[width=\columnwidth]{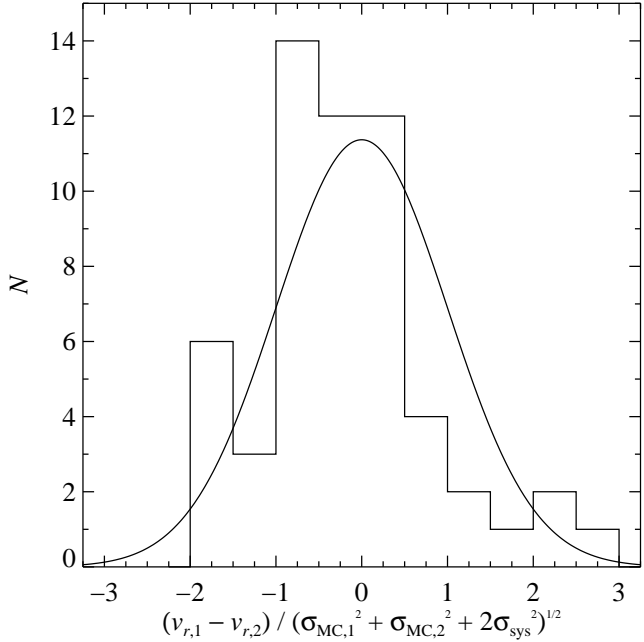}
\caption{Radial velocity measurement differences of the same stars
  observed on two different slitmasks.  The differences are divided by
  the quadrature sum of the measurement uncertainties, which include
  noise and systematic error terms.  The solid curve is a Gaussian
  with unit standard deviation.  The systematic error term,
  $\sigma_{\rm sys}$, was adjusted to ensure that the standard
  deviation is unity.\label{fig:vdiff}}
\end{figure}

\begin{figure}[t!]
\centering
\includegraphics[width=\columnwidth]{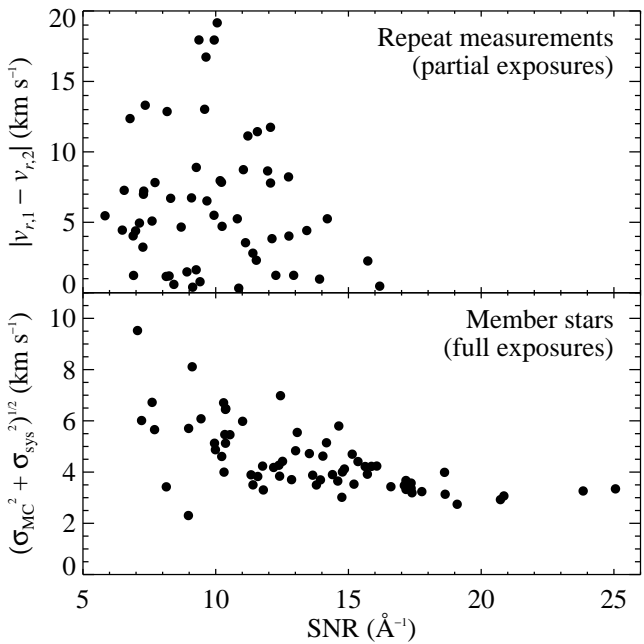}
\caption{Top: Absolute radial velocity differences of repeat
  measurements of the same stars on two different slitmasks versus
  SNR\@.  These measurements were necessarily based on spectra with
  about half the exposure time of the final, stacked spectra.  Bottom:
  Estimated error on $v_r$ as a function of SNR for all member stars.
  The total error is the quadrature sum of the random (Monte Carlo)
  and systematic error terms.\label{fig:vdiffsn}}
\end{figure}

From duplicate $v_r$ measurements of the same stars,
\citeauthor{sim07}\ realized that $\sigma_{\rm MC}$ was an incomplete
estimate of the radial velocity uncertainty.  In order to explain the
dispersion among the repeat measurements, they added a systematic
error term, $\sigma_{\rm sys}$, in quadrature with $\sigma_{\rm MC}$.
The systematic term was chosen such that the radial velocity
differences divided by their total uncertainties were distributed
normally with unit standard deviation.

\begin{equation}
\sqrt{\frac{1}{N} \sum_{i=1}^N \frac{\left(v_{r,1} - v_{r,2}\right)^2}{\sigma_{\rm MC,1}^2 + \sigma_{\rm MC,2}^2 + 2\sigma_{\rm sys}^2}} = 1
\label{eq:vsyserr}
\end{equation}

Because we observed 106 duplicate objects on both DEIMOS slitmasks, we
also calculated a systematic error term.  We excluded galaxies,
obvious foreground dwarfs with \ion{Na}{1}~8190 absorption, and other
objects failing any membership criterion other than radial velocity
(see Section~\ref{sec:membership}).  From the \nvdiff\ remaining
stars, we computed $\sigma_{\rm sys} = \vsyserr$~km~s$^{-1}$---which
is the same value that \citeauthor{sim07}\ derived---by solving
Equation~\ref{eq:vsyserr}.  Figure~\ref{fig:vdiff} shows the
distribution of differences in repeat $v_r$ measurements divided by
their estimated uncertainties.  The width of the distribution matches
a unit Gaussian because $\sigma_{\rm sys}$ was tuned to force such a
match.

Figure~\ref{fig:vdiffsn} shows how velocity errors and uncertainties
trend with SNR\@.  The $v_r$ differences between repeat measurements
(top panel) are based on exposures from one slitmask only.  Therefore,
their SNRs are about $1/\sqrt{2}$ of the SNR of the stacked spectra
used for remainder of our analysis.  The estimated uncertainties
(bottom panel) decrease with increasing SNR, as expected.  Only the
random error term, $\sigma_{\rm MC}$, decreases with SNR\@.  The
systematic error term, $\sigma_{\rm sys}$, is constant.

%%%%%%%%%%%%%%%%%%%%%%%%%%%%%%%%%
%%%%%%%%%   SECTION 4   %%%%%%%%%
%%%%%%%%%%%%%%%%%%%%%%%%%%%%%%%%%

\section{Membership}
\label{sec:membership}

We used several criteria for determining membership.  First, we
eliminated stars based on their positions in the CMD\@.  Then, we
eliminated stars with spectral features not found in red giants.
Finally, we eliminated stars based on their radial velocities.

\subsection{Color-Magnitude Diagram}
\label{sec:cmd}

\citeauthor*{bel11a}\ identified the extinction-corrected magnitude of
the TRGB as $r_0 = 22.61 \pm 0.06$.  In selecting spectroscopic
targets, we prioritized stars with $r > 22$, which conservatively
allowed for large error in the TRGB magnitude.  It also allowed some
asymptotic giant branch (AGB) stars.  In order to fill the slitmasks,
we also included some targets with $r \le 22$.

Stars brighter than the TRGB are not likely members.  We carefully
examined all stars with $r_0 < 22.49$ ($2\sigma$ brighter than the
TRGB).  After excluding stars failing to meet the membership criteria
in the following sections, four stars remained.  Star 8116 showed
broad CaT lines.  The width and shape suggested pressure broadening
typical of dwarf stars.  Therefore, we ruled 8116 as a non-member.
The spectrum of star 8445 looked in all respects like a red giant.
Although the star is slightly brighter and bluer than the TRGB, we
kept 8445 in the member list under the premise that it is an AGB star.
The spectrum of star 11163 showed very strong cyanogen bands.  If this
carbon star is an AGB member of VV124, it could be significantly
brighter and redder than the TRGB\@.  In fact, that is its location in
the CMD\@.  For this reason, we kept 11163 in the membership list.
Finally, star 13649 is solidly in the ``wall'' of foreground dwarfs in
the CMD, and its \ion{Na}{1}~8190 equivalent width (EW) is very close
to the cut-off for membership (see the next section).  We ruled 13649
a foreground dwarf for the combination of its position in the CMD and
moderately strong \ion{Na}{1}~8190 doublet.

\subsection{Spectral Features}

Inspection of the DEIMOS spectra revealed that many of our targets
were background galaxies or foreground dwarf stars.  First, we
excluded all objects with emission lines or broad, redshifted Ca~H and
Ca~K in absorption.  These 20 objects are background galaxies.

Second, we excluded stars with very strong, broad \ion{Na}{1}~8190
doublets.  The strengths of these lines are especially sensitive to
surface gravity.  That sensitivity makes them excellent discriminators
between giants and foreground dwarf stars
\citep{spi71,coh78,sch97,gil06}.  We measured the EWs of each of the
two lines in the doublet by fitting two independent Gaussians.  We
computed Monte Carlo uncertainties by resampling the spectrum 100
times.  For each pixel in each resample, we added noise from a
Gaussian random distribution with a standard deviation equal to the
estimated Poisson noise of that pixel.  We fitted Gaussians to each
resample, and we adopted the standard deviation among their EWs as the
error on the EW\@.  Table~\ref{tab:catalog} gives the summed EWs and
their uncertainties for stars where the doublet was measurable above
the noise.

\begin{figure}[t!]
\centering \includegraphics[width=\columnwidth]{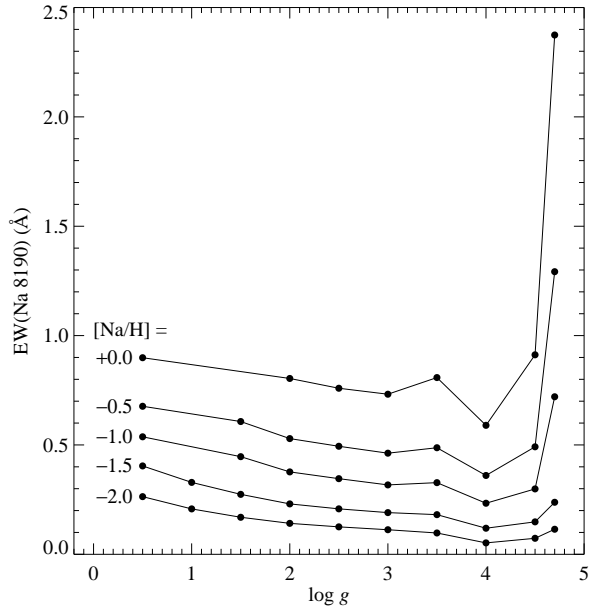}
\caption{Summed EW of the \ion{Na}{1}~8190 doublet as a function of
  surface gravity and [Na/H]\@.  The summed EW exceeds 1~\AA\ only for
  dwarf stars ($\log g > 4.5$).\label{fig:nlte}}
\end{figure}

We computed a threshold value for the summed EW of the doublet by
looking up EWs in a grid of non-local thermodynamic equilibrium (NLTE)
Na abundances \citep{lin11}.  We sampled EWs from $\log g = 0.5$ to
$4.7$ and ${\rm [Na/H]} = 0.0$ to $-2.0$.  We found the appropriate
effective temperature for each combination of surface gravity and
abundance (assuming ${\rm [Na/Fe]} = 0$) from a 10~Gyr Yonsei-Yale
isochrone \citep{dem04}.  Interpolations in the NLTE grid yielded
Figure~\ref{fig:nlte}.  Some points are missing from the figure
because they are also missing from the NLTE grid.  Regardless, it is
clear that the doublet increases sharply in EW for dwarf stars with
$\log g > 4$.  The summed EW of the doublet exceeds 1~\AA\ for stars
with $\rm{[Na/H]} \le 0$ only at surface gravities $\log g > 4.5$.  We
identified 17 stars with a summed \ion{Na}{1}~8190 EW greater than
1~\AA\@.  These 17 stars are foreground dwarfs.

%We computed synthetic EWs in order to determine a threshold value
%above which the star should be classified as a foreground dwarf.  We
%used the spectral synthesis code MOOG \citep{sne73}, ATLAS9 model
%atmospheres in local thermodynamic equilibrium \citep{kur93,cas04},
%and oscillator strengths from the online NIST database
%\citep{ral11,mei37}.  We found that the summed EW of the doublet
%exceeds 0.8~\AA\ for stars with $\rm{[Na/H]} \le 0$ only at surface
%gravities $\log g > 3$.  The summed EW of the typical member star in
%our sample should be 0.4~\AA\@.  We identified 18 stars with a summed
%\ion{Na}{1}~8190 EW greater than 0.8~\AA\@.  These 18 stars are
%foreground dwarfs.

Counting star 8116, excluded for its broad calcium lines and for being
brighter than the TRGB, we eliminated 38 objects based on spectral
features.  Another three stars showed strong CN bands.  One of these,
11163, is discussed in Section~\ref{sec:cmd}.  The other two stars,
9378 and 13043, pass all membership criteria, and we retained them as
members.

%Some spectra showed additional features that do not necessarily
%indicate non-membership.  Twenty stars exhibited obvious, deep TiO
%bands.  When these bands are seen in stars with $g-r < 1.5$, the stars
%are usually dwarfs.  However, cool, metal-rich giants can also have
%strong TiO\@.  Regardless, all of the TiO stars were eliminated for
%other reasons.

\subsection{Radial Velocities}

\begin{figure}[t!]
\centering \includegraphics[width=\columnwidth]{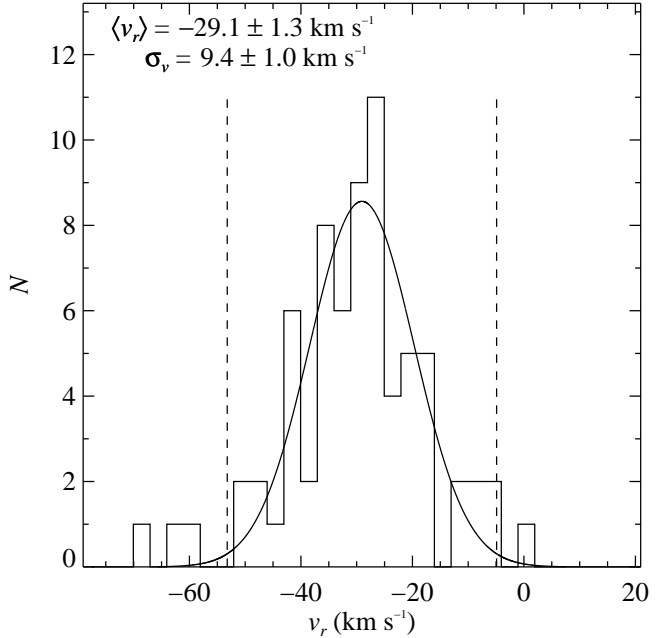}
\caption{Heliocentric radial velocity distribution.  The solid curve
  is a Gaussian distribution corresponding to the measured mean
  velocity and velocity dispersion.  The dashed lines enclose 99\%
  ($\pm 2.58\sigma$) of the member stars.  The figure label gives the
  average heliocentric velocity and the velocity dispersion corrected
  for measurement uncertainty (see Section~\ref{sec:dynamics}).  Only
  stars passing all membership cuts except velocity are
  shown.\label{fig:vhist}}
\end{figure}

\addtocounter{table}{1}

The median radial velocity uncertainty in our sample of member stars
is $\medianvrerr$~km~s$^{-1}$, on the order of the velocity dispersion
of a small dSph.  Therefore, it is important to consider velocity
errors carefully when computing the velocity dispersion.
\citet{wal06} devised a formalism to take these errors into account.
The procedure involved maximizing the likelihood that the average
radial velocity, $\langle v_r \rangle$, and the intrinsic velocity
dispersion, $\sigma_v$, accurately describe the observed velocity
distribution of candidate member stars in the presence of
observational error.  The process was iterative.  We chose $\langle
v_r \rangle = -30$~km~s$^{-1}$ and $\sigma_v = 9$~km~s$^{-1}$ as rough
guesses for the first iteration.  In each subsequent iteration, only
stars within $2.58\sigma_v$ (99\% of a normal distribution) of
$\langle v_r \rangle$ were considered.  In fact, just one iteration
was required to converge on the final member sample and the final
values of $\langle v_r \rangle$ and $\sigma_v$.  The uncertainties on
$\langle v_r \rangle$ and $\sigma_v$ were determined from the
likelihood covariance matrix.  Table~\ref{tab:parameters} gives the
fiducial values for $\langle v_r \rangle$ and $\sigma_v$ and their
uncertainties.

We used $\langle v_r \rangle$ and $\sigma_v$ to determine the final
membership cut.  If a normal distribution is a good description of the
velocity distribution, then the velocity range $\langle v_r \rangle
\pm 2.58 \sigma_v$ includes 99\% of the members.  This cut excluded
\nvexclude\ stars and retained \nmember\ members.  Some galaxy
dynamics studies suggest that a cut of $3\sigma$ is a better choice.
Adopting a $3\sigma$ cut for VV124 would not have affected the
membership of any star.

Table~\ref{tab:catalog} lists all of the spectroscopic targets.  The
table gives astrometric and photometric data, as well as quantities
derived from the spectra.  The last two columns of the table indicate
whether the object is a member star and the reasons that stars were
considered non-members.

%Adopting a $3\sigma$ cut for VV124 would have included one additional
%star, 5248.  At 3.2', this star is one of the farthest from the center
%of VV124\@.  It could be a star revolving in the disk of VV124, or it
%could be a non-member.  We ruled it a non-member for two reasons.
%First, the five member stars nearest to 5248 have $v_r$ within
%5~km~s$^{-1}$ of $\langle v_r \rangle$.  They clearly do not belong to
%an edge-on rotating disk.  Second, the galaxy shows no evidence of
%rotation (see Section~\ref{sec:rotation}).  Consequently, there is no
%reason to prefer a $3\sigma$ velocity membership cut over $2.58\sigma$
%(99\%).  We used $2.58\sigma$.

Figure~\ref{fig:vhist} presents the velocity distribution of VV124\@.
The velocities are distributed normally.  The skewness, a measure of
asymmetry, of the $v_r$ distribution is $\vskewness \pm
\vskewnesserr$.  The kurtosis, a measure of the deviation from
Gaussianity, is $\vkurtosis \pm \vkurtosiserr$.  Both parameters are
consistent with zero, indicating conformance to a normal distribution.
The vertical, dashed lines show $\langle v_r \rangle \pm
2.58\sigma_v$, and the solid curve shows a perfect normal distribution
(not a fit) for those parameters, normalized to the number of member
stars.

%%%%%%%%%%%%%%%%%%%%%%%%%%%%%%%%%
%%%%%%%%%   SECTION 5   %%%%%%%%%
%%%%%%%%%%%%%%%%%%%%%%%%%%%%%%%%%

\section{Dynamics}
\label{sec:dynamics}

\begin{deluxetable*}{lcc}
\tablewidth{0pt}
\tablecolumns{3}
\tablecaption{Properties of VV124\label{tab:parameters}}
\tablehead{\colhead{Property} & \colhead{Symbol} & \colhead{Value}}
\startdata
\cutinhead{Photometry}
Distance\tablenotemark{a}                    & $D$                            & $1.36 \pm 0.03$~Mpc \\
Luminosity\tablenotemark{b}                  & $L_V$                          & $8.2^{+1.6}_{-1.4} \times 10^6~L_{\sun}$ \\
Stellar mass\tablenotemark{c}                & $M_*$                          & $9.4^{+3.8}_{-2.9} \times 10^6~M_{\sun}$ \\
Gas mass\tablenotemark{b}                    & $M_{\rm gas}$                  & $8.7 \times 10^5~M_{\sun}$ \\
Half-light radius\tablenotemark{b}           & $R_e$                          & $41.3'' = 260$~pc \\
\cutinhead{Dynamics}
Mean radial velocity                         & $\langle v_r \rangle$          & $-29.1 \pm 1.3$~km~s$^{-1}$ \\
Line-of-sight velocity dispersion            & $\sigma_v$                     & $ 9.4 \pm 1.0$~km~s$^{-1}$ \\
Mass within half-light radius\tablenotemark{d} & $M_{1/2}$                    & $(2.12 \pm 0.22) \times 10^7~M_{\sun}$ \\
Mass-to-light ratio within half-light radius\tablenotemark{d} & $(M/L_V)_{1/2}$ & $5.2 \pm 1.1~M_{\sun}/L_{\sun}$ \\
Dynamical-to-stellar mass ratio within half-light radius\tablenotemark{c,d} & $(M_{\rm dyn}/M_*)_{1/2}$ & $4.5 \pm 1.9$ \\
Total mass\tablenotemark{e}                  & $M_{\rm tot}$                  & $(1.95 \pm 0.40) \times 10^7~M_{\sun}$ \\
Total mass-to-light ratio\tablenotemark{e}   & $(M/L_V)_{\rm tot}$              & $4.8 \pm 1.3~M_{\sun}/L_{\sun}$ \\
\cutinhead{Metallicity}
Mean metallicity\tablenotemark{f}            & $\langle {\rm [Fe/H]} \rangle$ & $-1.58 \pm 0.06$ \\
Standard deviation                           & $\sigma({\rm [Fe/H]})$         & $0.51$ \\
Median metallicity                           & med([Fe/H])                    & $-1.52$ \\
Median absolute deviation                    & mad([Fe/H])                    & $0.27$ \\
Interquartile range                          & IQR([Fe/H])                    & $0.58$ \\
Skewness                                     & Skew([Fe/H])                   & $-0.51 \pm 0.31$ \\
Kurtosis                                     & Kurt([Fe/H])                   & $-0.01 \pm 0.60$ \\
Yield (Simple Model)                         & $p$ (Simple)                   & $0.045^{+0.007}_{-0.006}~Z_{\sun}$ \\
Yield (Pre-Enriched Model)                   & $p$ (Pre-Enriched)             & $0.044^{+0.008}_{-0.007}~Z_{\sun}$ \\
Initial metallicity (Pre-Enriched Model)     & ${\rm [Fe/H]}_0$               & $< -3.21$ \\
Yield (Extra Gas Model)                      & $p$ (Extra~Gas)                & $0.041^{+0.006}_{-0.005}~Z_{\sun}$ \\
Extra Gas Parameter (Extra Gas Model)        & $M$                            & $2.86^{+2.01}_{-1.15}$ \\
\enddata
\tablerefs{a: \protect \citet{jac11}.  b: \protect \citeauthor*{bel11a}.}
\tablenotetext{c}{Although \protect \citeauthor*{bel11a} assumed $M_*/L_V = 2$, we assumed that $M_*/L_V = 1.10$, a value more typical of transition-type dwarfs \protect \citep{woo08}.  We used this stellar mass-to-light ratio to calculate these quantities.  The value could range from 0.8 to 1.5, and we took these limits into account in the estimation of the uncertainties.}
\tablenotetext{d}{Using the formula $M_{1/2} = 4 G^{-1} R_e \sigma_v^2$ \protect \citep{wol10}.}
\tablenotetext{e}{Using the formula $M_{\rm tot} = 167 \mu r_c \sigma_v^2$ \protect \citep{ill76}, which may not be appropriate for VV124\@.  This formula often gives smaller $M_{\rm tot}$ than the formula for $M_{1/2}$.}
\tablenotetext{f}{Although \protect \citet{kir11a} calculated $\langle {\rm [Fe/H]} \rangle$ with inverse variance weighting, we present the unweighted mean.  The low SNRs of many of our spectra cause the error on [Fe/H] to increase strongly with decreasing metallicity.  Therefore, the weighted mean would be biased toward high [Fe/H].}
\end{deluxetable*}

\subsection{Mean Radial Velocity and Velocity Dispersion}
\label{sec:vdisp}

The mean radial velocity, $\langle v_r \rangle = \meanv \pm
\meanverr$~km~s$^{-1}$, agrees with the radial velocity of the
\ion{H}{1} gas, $\langle v_r \rangle = -25 \pm 4$~km~s$^{-1}$
(\citeauthor*{bel11a}).  The agreement is especially reassuring
because previous optical measurements of radial velocities showed
significantly more negative $v_r$.  \citet{kop08} and \citet{tik10}
observed two blue supergiants: bl1 with $v_r = -90 \pm 15$~km~s$^{-1}$
and bl2 with $v_r = -82 \pm 15$~km~s$^{-1}$.  They also estimated the
average radial velocity of unresolved stellar light to be $\langle v_r
\rangle = -70 \pm 15$~km~s$^{-1}$.  Radial velocities of resolved and
diffuse \ion{H}{2} gas, measured from the [\ion{O}{3}]~5007 emission
line, ranged from $-36$ to $-71$~km~s$^{-1}$.
\citeauthor*{bel11a}\ revised the velocity of the supergiant bl2 to
$v_r = -44 \pm 18$~km~s$^{-1}$.  They also observed a star cluster,
C1, with $v_r = -86 \pm 20$~km~s$^{-1}$ and diffuse H$\alpha$ emission
with $\langle v_r \rangle = -6 \pm 30$~km~s$^{-1}$.  The individual
sources bl1, C1, and one of the \ion{H}{2} regions have radial
velocities more than $1\sigma$ inconsistent with our velocity cuts for
membership.  The stellar integrated light measurement of $\langle v_r
\rangle$ by \citeauthor{kop08}\ and \citeauthor{tik10}\ is $2.7\sigma$
discrepant from our measurement.  In every case, the discrepancies are
such that the other measurements are more negative than ours.  We are
confident that our measurement of $\langle v_r \rangle$ is accurate
because it agrees with the \ion{H}{1} measurement.  Moreover, our
estimate is based on higher-resolution spectra and a much larger
sample than \citet{kop08}, \citet{tik10}, and \citeauthor*{bel11a}.

There are two possible causes of the discrepancy.  First, there could
have been some systematic error in the measurements of $v_r$ from the
low-resolution spectra of \citeauthor{kop08}\ and \citeauthor{tik10}
The upward revision of $v_r$ for bl2 by \citeauthor*{bel11a}\ may
support this conclusion.  On the other hand,
\citeauthor*{bel11a}\ also measured a very negative $v_r$ for the star
cluster C1.  We also note that we excluded three stars with $-75~{\rm
  km~s}^{-1} < v_r < -55~{\rm km~s}^{-1}$ as non-members because their
velocities are too negative.  These velocities are similar to those of
bl1 and C1.  It may also be relevant that the \ion{H}{1} gas southeast
of the center of the galaxy has more negative $v_r$ than the gas in
the central regions (\citeauthor*{bel11a}).  Thus, VV124 may have a
non-equilibrium dynamical component consisting of a young stars and
star-forming regions.  This component would have more negative $v_r$
than the old red giants and neutral gas.  The data available do not
point to a definitive conclusion.  Higher resolution observations of
bl1, C1, other young stars, and star-forming regions will better
define VV124's velocity structure.

Unresolved binaries could inflate the velocity dispersion.
\citet{min10} concluded that this inflation is unlikely to be more
than 30\% in galaxies with $4~{\rm km~s}^{-1} < \sigma_v < 10~{\rm
  km~s}^{-1}$.  In a separate study, \citet{mcc10} arrived at a
similar conclusion, but cautioned that even a modest binary fraction
could cause an ultra-faint dSph-sized stellar population free of dark
matter to appear to have the velocity dispersion of a dark
matter-dominated system.  In a confirmation of the contribution of
unresolved binaries to mass estimates of dSphs, \citet{kop11} revised
the velocity dispersion of the Bo{\"o}tes~I dSph from $>6$~km~s$^{-1}$
\citep{mun06,mar07} to two components: one population at
2.4~km~s$^{-1}$ and a smaller population at $\sim 9$~km~s$^{-1}$.
Most of this revision was due to a more precise estimation of velocity
uncertainties, but removing binaries did contribute to the reduction
in $\sigma_v$.

We did not obtain multi-epoch spectroscopy for VV124\@.  Therefore, we
were unable to remove the contribution of unresolved binaries.  We
suggest that the binary contribution to $\sigma_v$ for VV124 is not
nearly as large as for Bo{\"o}tes~I or other ultra-faint dSphs.  The
warnings about the inflation of velocity dispersion by \citet{min10},
\citet{mcc10}, and \citet{kop11} specifically concerned the
ultra-faint dSphs with very low luminosities ($<10^5~L_{\sun}$).
VV124 does not fall into this class.  \citet{ols96} showed that
unresolved binaries inflate $\sigma_v$ for the Ursa Minor and Draco
dSphs by much less than 1~km~s$^{-1}$.  Because the velocity
dispersion of VV124 is closer to the dispersions of Ursa Minor and
Draco than to ultra-faint dSphs, binarity likely does not dominate our
measurement uncertainty.

\subsection{Rotation}
\label{sec:rotation}

\begin{figure}[t!]
\centering
\includegraphics[width=\columnwidth]{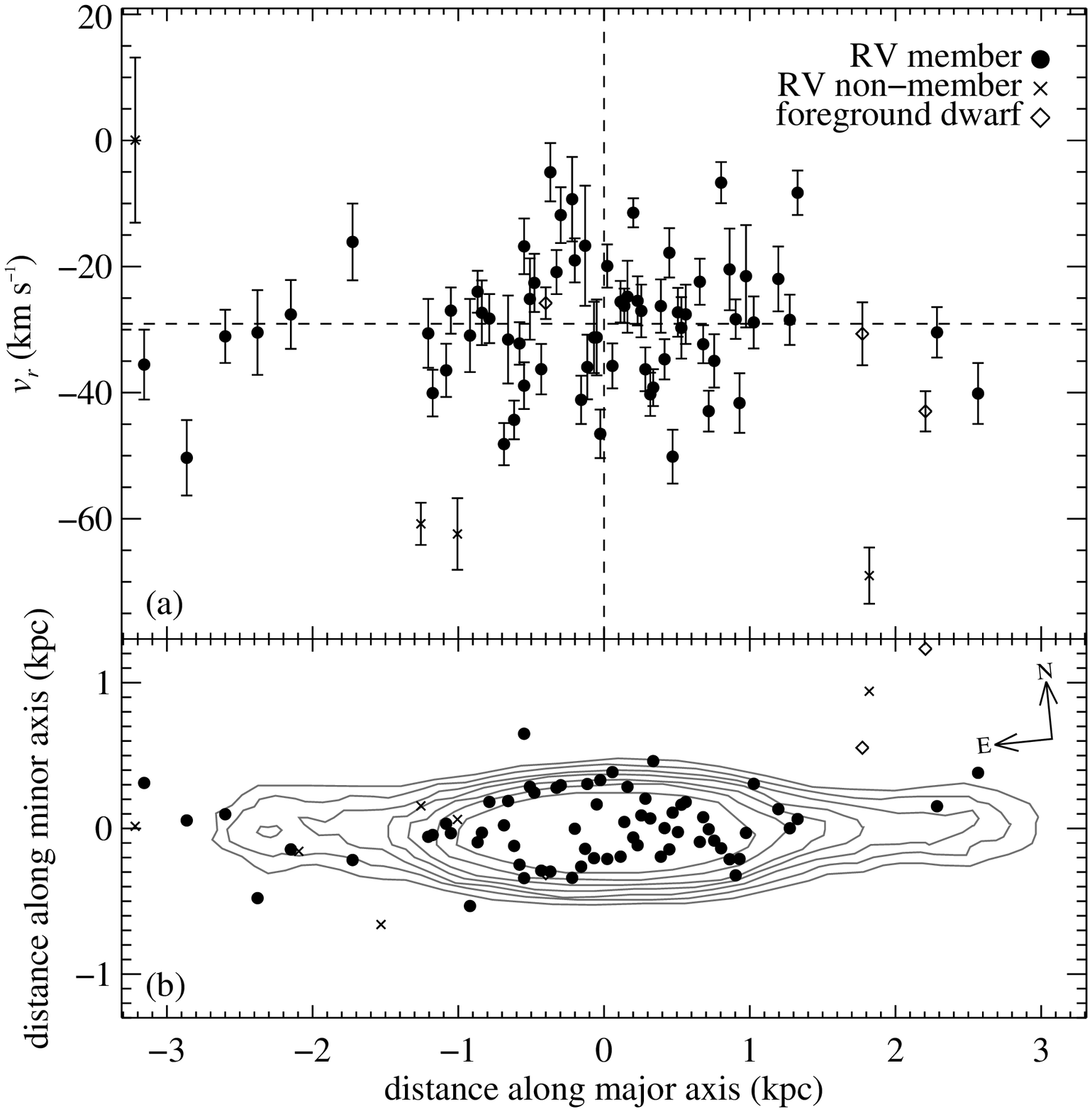}
\caption{(a) Heliocentric radial velocity as a function of
  displacement from the minor axis (distance along the major axis).
  Filled points are spectroscopic members, and hollow diamonds and
  crosses are non-members.  The horizontal dashed line indicates
  $\langle v_r \rangle$.  Only stars passing all membership cuts
  except velocity are shown.  (b) Spatial locations of members and
  stellar non-members fainter than the TRGB\@.  The gray lines are
  contours of equal surface density (\protect\citeauthor*{bel11a}).
  From the inside to the outside, the contours represent $3\sigma$,
  $4\sigma$, $5\sigma$, $6\sigma$, $7\sigma$, $8\sigma$, and $9\sigma$
  differences from the central surface density.\label{fig:rot}}
\end{figure}

While \citeauthor*{bel11a}\ concluded that the \ion{H}{1} gas in VV124
does not show any signature of rotation, the interpretation of the
stellar wings as an edge-on disk of stars prompted us to search for
stellar rotation.  If the structure is a disk, it is nearly edge-on,
and rotation in principle should be evident in measurements of stellar
radial velocities.

VV124 also contains \ion{H}{1} gas that has a velocity gradient of
5--10~km~s$^{-1}$ across the face of the galaxy and a full velocity
range of $\sim 45$~km~s$^{-1}$.  However, the gradient is not aligned
with the disk-like structure.  \citeauthor*{bel11a}\ suggested that
the gradient arises not from rotation, but asymmetric stellar winds
acting preferentially on the southeast quadrant of the galaxy.

Based on \citeauthor*{bel11a}'s Figure~11 and our
Figure~\ref{fig:rot}, we define the ``wings'' of VV124 as $|X| \ge 4'$
(1.6~kpc) , where $X$ is distance along the major axis.  Our sample
includes eight such member stars.  Therefore, we have
spectroscopically confirmed the existence of the photometrically
feeble disk-like structure discovered by \citeauthor*{bel11a}.

However, these stars show no evidence of rotation within the
limitations of our measurements.  Figure~\ref{fig:rot} shows stellar
radial velocities versus distance from the minor axis.  We quantified
the amount of rotation as $\Delta v_{\rm EW}$, the difference between
the average $v_r$ in the eastern and western wings.  Our sample
contains six members in the eastern wing and two members in the
western wing.  The velocity measurements of these stars yielded
$\Delta v_{\rm EW} = \avgvdiff$~km~s$^{-1}$.

In order to assess the precision of this measurement, we constructed
theoretical rotation curves for the outermost regions of the galaxy.
The curves were flat in the wings ($r > 1.5'$), and their magnitudes
ranged from 0 to 20~km~s$^{-1}$, spaced at 1~km~s$^{-1}$.  We sampled
the theoretical rotation curves with eight fake stars at the same
distances along the major axis as the stars in our sample.  Each fake
star had a radial velocity perturbed from the rotation curve, and the
amount of perturbation was sampled from a Gaussian random probability
distribution with a standard deviation equal to the estimated
uncertainty in $v_r$ for the corresponding observed star.  Under the
assumption that any rotation would be diluted by incoherent velocity
dispersion, we further perturbed each star's velocity according to a
Gaussian random distribution with a standard deviation of $\sigma_v$,
the previously measured line-of-sight velocity dispersion of VV124\@.
We computed $\Delta v_{\rm EW}$ for 1000 such Monte Carlo trials for
each of the rotation curves.  The standard deviation of the
probability distributions is an estimate of the uncertainty on the
observed value of $\Delta v_{\rm EW} = \avgvdiff$~km~s$^{-1}$.  This
uncertainty, $\avgvdifferr$~km~s$^{-1}$, is insensitive to the
magnitude of the rotation.

Our measurements are consistent with the absence of rotation because
$\Delta v_{\rm EW}$ is consistent with zero.  However, our data set is
limited in its ability to detect rotation.  Under the assumption that
the velocity dispersion is constant with radius and that this velocity
dispersion dilutes any rotation signal in the wings, the minimum
rotation signal we could have detected is $\avgvdifferr$~km~s$^{-1}$.

%The maximum mass enclosed within the wings (roughly within 2~kpc) is
%$M(< 2~{\rm kpc}) = (v_{\rm rot}^2 + 3\sigma_v^2)R/G = \maxmassinwings
%\times 10^8~M_{\sun}$.

%Rotation would manifest in coherent velocities different from $\langle
%v_r \rangle$.  Based on our measurement of dynamical mass
%(Section~\ref{sec:mass}), the circular velocity in the wings would be
%$v_{\rm circ} = GM(<R)/R = \vrotlo$--$\vrothi$~km~s$^{-1}$, where we
%have approximated $M(<R)$, the enclosed mass, as $M_{1/2}$, the mass
%within the half-light radius.  The dotted curves in
%Figure~\ref{fig:rot} delineate $v_{\rm circ}$ based on that mass.  The
%circular velocity is a crude approximation to the rotational velocity.
%The precise magnitude and shape of the rotation curve depends on (a)
%the ratio of rotational and dispersion support and (b) the shape of
%the mass profile.  Our assumption that $v_{\rm rot} = v_{\rm circ}$
%overestimates the rotational velocity, and our assumption that the
%mass enclosed is $M_{1/2}$ underestimates the rotational velocity.
%The present data set constrains neither $v_{\rm rot}/\sigma_v$ nor the
%mass profile.  Therefore, we base the following discussion on our
%crude assumptions.

The stars in the southeast of the galaxy do not have any more negative
$v_r$ than stars elsewhere.  Therefore, our observations do not show a
stellar counterpart to the asymmetric gas velocity.  Radial velocity
measurements of bluer, younger stars might show such structure, but we
have not performed such observations.

Our observations are not conclusive evidence against rotation or a
stellar counterpart to the \ion{H}{1} velocity structure.
Furthermore, we may have misclassified some foreground stars as
members.  Such misclassifications may alter a rotation signal.  We
considered the possibility that some true members may have been
misclassified as foreground dwarfs based on the strength of their
\ion{Na}{1}~8190 doublets (hollow diamonds in Figure~\ref{fig:rot}),
but the radial velocities of those stars do not suggest rotation or
velocity structure any more than the spectroscopic members.  We cannot
rule out the possibility that a larger sample with more precise
measurements of $v_r$ will detect velocity structure in the stars.

\subsection{Mass}
\label{sec:mass}

The mass of a spheroidal stellar system can be determined from its
line-of-sight velocity dispersion under the assumption of a mass
profile and velocity anisotropy.  In the case of VV124, only stars and
dark matter significantly affect $\sigma_v$.  Gas is irrelevant
because the gas mass is just a small fraction of the stellar mass.
\citet{ill76} presented a formula to measure the mass of globular
clusters: $M_{\rm tot} = 167 \mu r_c \sigma_v^2$, where $r_c$ is the
\citet{kin62} core radius and $\mu$\footnote{\citet{kin66} and
  \citet{ill76} called this parameter $\mu$.  \citet{mat98} and
  \citet{sim07} called it $\beta$.  Because $\beta$ is typically used
  to represent velocity dispersion anisotropy, we have chosen to call
  the concentration parameter $\mu$.} is a dimensionless number that
quantifies the mass concentration toward the system's center.  The
core radius may be approximated in relation to $R_e$, the half-light
or effective radius, as follows: $r_c = 0.64R_e$.  \citet{mat98}
stated that $\mu = 8$ is appropriate for low-concentration King
profiles and therefore appropriate for most dwarf galaxies.  However,
\citeauthor{ill76}'s (\citeyear{ill76}) formula assumes that mass
follows light and that the velocity dispersion anisotropy $\beta = 0$.
These two assumptions are appropriate for globular clusters but not
for dark matter-dominated galaxies.  \citet{wol10} presented an
alternative formula for the mass enclosed within the half-light
radius: $M_{1/2} = 4 G^{-1} R_e \sigma_v^2$, where $R_e$ is the
half-light or effective radius.  It is at this radius that $\beta$ has
the least influence on the estimate of the enclosed mass.  Inside and
beyond $R_e$, the enclosed mass becomes increasingly uncertain because
$\beta$ is unknown when only line-of-sight velocities are known.
Table~\ref{tab:parameters} gives both $M_{\rm tot}$ from the
\citet{ill76} formula and $M_{1/2}$ from the \citet{wol10} formula.
As is typical for dwarf galaxies, $M_{1/2}$ actually exceeds $M_{\rm
  tot}$ because the \citet{ill76} formula is inappropriate for dwarf
galaxies.  Note that both of these formulae assume sphericity.  The
ellipticity of VV124 ($\epsilon = 0.44$, \citeauthor*{bel11a})
diminishes the accuracy of these mass estimators for VV124\@.

The mass-to-light ratio within the half-light radius is $(M/L_V)_{1/2}
= \ml \pm \mlerr~M_{\sun}/L_{\sun}$.  This value is too large for a
mass dominated by stars alone.  According to up-to-date stellar models
\citep{per09}, old (12~Gyr) and metal-poor (${\rm [Fe/H]} = -1$, which
is about the mean metallicity for VV124, as derived in
Section~\ref{sec:metallicity}) stellar populations without dark matter
exhibit $M/L_V \le 2.5~M_{\sun}/L_{\sun}$.  Due to the contribution of
the young population, the expected value for VV124 should be lower.
The ratio of the dynamical mass to the stellar mass within the
half-light radius is $(M_{\rm dyn}/M_*)_{1/2} = \mratio \pm
\mratioerr$.  (This value depends inversely on the assumed stellar
mass-to-light ratio, $M_*/L_V$.  See footnote c in
Table~\ref{tab:parameters}.)  Therefore, we conclude that the mass of
VV124 is dominated by dark matter.

\citeauthor*{bel11a}\ roughly estimated the mass-to-light ratio
($M/L_V \approx 8$) of VV124 from the \ion{H}{1} velocity
distribution.  They assumed that the gravitational potential dominates
the gas motion and that the \ion{H}{1} distribution is spherical and
isotropic.  Their value is \mlratio\ times our measurement of
$(M/L_V)_{1/2}$.  However, theirs was a rough estimate.  The
difference can be explained by their slightly larger measurement of
11~km~s$^{-1}$ for the velocity dispersion, compared to our
measurement of $\sigmav$~km~s$^{-1}$.  Our measurement is more
accurate because it is based on individual stars.  The gas velocity
dispersion could be sensitive to stellar feedback, especially in light
of \citeauthor{bel11a}'s interpretation of stellar winds as the cause
of the velocity gradient seen in the gas.

%%%%%%%%%%%%%%%%%%%%%%%%%%%%%%%%%
%%%%%%%%%   SECTION 6   %%%%%%%%%
%%%%%%%%%%%%%%%%%%%%%%%%%%%%%%%%%

\section{Metallicity Distribution}
\label{sec:metallicity}

\begin{figure}[t!]
\centering
\includegraphics[width=\columnwidth]{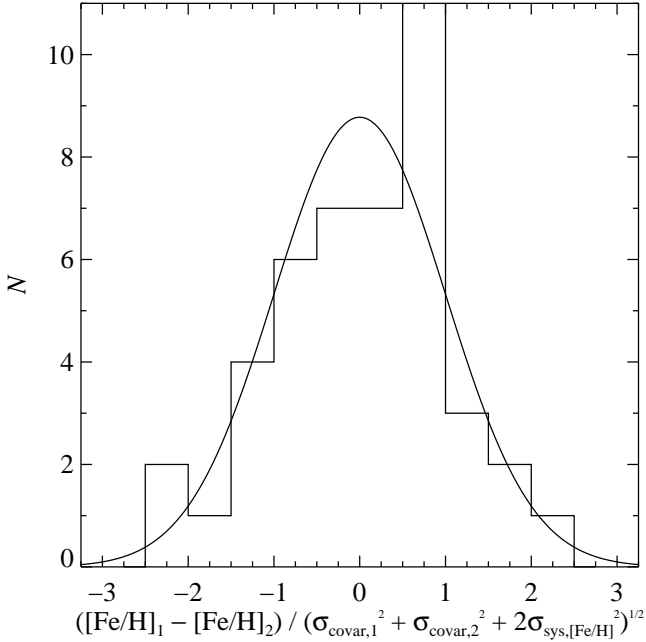}
\caption{Metallicity measurement differences of the same stars
  observed on two different slitmasks.  The differences are divided by
  the quadrature sum of the measurement uncertainties, which include
  noise and systematic error terms.  The solid curve is a Gaussian
  with unit standard deviation.  The systematic error term,
  $\sigma_{\rm sys,[Fe/H]}$, was adjusted to ensure that the standard
  deviation is unity.  Compare this figure to
  Figure~\ref{fig:vdiff}.\label{fig:fehdiffhist}}
\end{figure}

\begin{figure}[t!]
\centering
\includegraphics[width=\columnwidth]{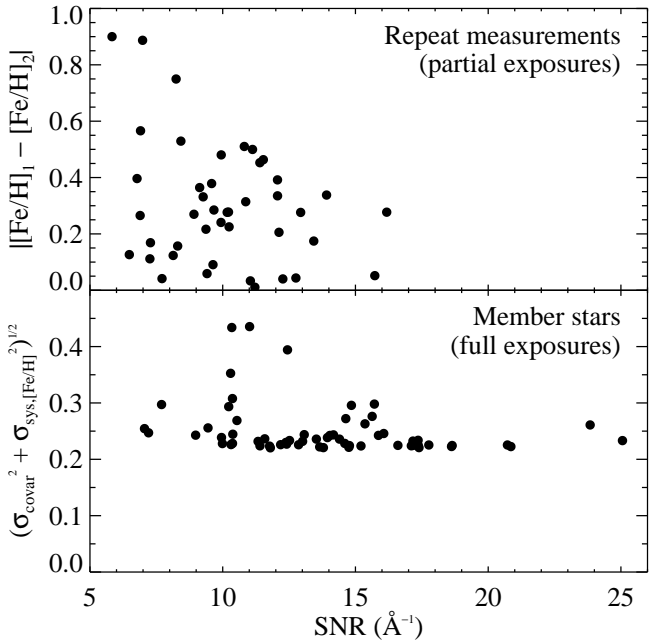}
\caption{Top: Absolute metallicity differences of repeat measurements
  of the same stars on two different slitmasks versus SNR\@.  These
  measurements were necessarily based on spectra with about half the
  exposure time of the final, stacked spectra.  Bottom: Estimated
  error on [Fe/H] as a function of SNR for all member stars.  The
  total error is the quadrature sum of the random and systematic error
  terms.  Compare this figure to
  Figure~\ref{fig:vdiffsn}.\label{fig:fehdiffsn}}
\end{figure}

We measured iron abundances from \ion{Fe}{1} lines in the DEIMOS
spectra.  \citet{kir08,kir09,kir10} described the procedure that we
used for these measurements.  We corrected the photometry for
extinction with \citeauthor{sch98}'s (\citeyear{sch98}) dust maps.
The average reddening was low ($E(B-V) \approx 0.01$).  After we
converted the extinction-corrected apparent magnitudes to absolute
magnitudes assuming $(m-M)_0 = 25.67 \pm 0.04$ \citep{jac11}, we used
12~Gyr Padova isochrones \citep{gir02} to estimate the photometric
effective temperature and surface gravity of each star.  A large grid
of synthetic spectra was searched for the best-fitting synthetic
spectrum.  In the search, temperature and [Fe/H] were varied.  The
temperature was assigned a probability distribution based on the
uncertainty in the star's position in the CMD\@.  This probability
distribution was included in the $\chi^2$ minimization to find the
best-fitting synthetic spectrum.  In this way, the spectroscopic
temperature was not allowed to stray far from the photometric
temperature.  (See Sections 4.5 and 4.7 of \citeauthor{kir09}
\citeyear{kir09} for more details.)  The photometry and the
spectroscopy shared roughly equal weight in determining the final
temperature.  The photometry alone determined the surface gravity.

In previous works based on this technique, the uncertainty in [Fe/H]
($\sigma_{\rm covar}$) was calculated in part from the covariance
matrix in the Levenberg-Marquardt optimization.  Various diagnostics
indicated that this random uncertainty term underestimated the full
error in the measurement.  Therefore, a systematic error term,
$\sigma_{\rm{sys,[Fe/H]}} = 0.113$, was added in quadrature to the
random uncertainty.  That term was calculated from spectra with a
typical exposure time of one-hour.  The exposure time for most spectra
in this work exceeds seven hours.  Therefore, errors in sky
subtraction play a much larger role than in our previous works.

Because the spectral noise behavior assumes a different character than
in the spectra from which the systematic error term was originally
calculated, we re-evaluated the magnitude of
$\sigma_{\rm{sys,[Fe/H]}}$ using only the VV124 data set.  In analogy
to our calculation of radial velocity uncertainties, we examined
differences in [Fe/H] between measurements of the same stars on the
two separate slitmasks.  We determined $\sigma_{\rm{sys,[Fe/H]}} =
0.216$ from the following equation.

\begin{equation}
\sqrt{\frac{1}{N} \sum_{i=1}^N \frac{\left(\rm{[Fe/H]}_{1} - \rm{[Fe/H]}_{2}\right)^2}{\sigma_{\rm covar,1}^2 + \sigma_{\rm covar,2}^2 + 2\sigma_{\rm sys,\rm{[Fe/H]}}^2}} = 1
\label{eq:fehsyserr}
\end{equation}

\noindent
In other words, $\sigma_{\rm{sys,[Fe/H]}} = 0.216$ is the value
required to force $1\sigma$ agreement between the repeat measurements
of [Fe/H]\@.  Figure~\ref{fig:fehdiffhist} shows that this agreement
was achieved.

Figure~\ref{fig:fehdiffsn} shows the relationship between the error or
uncertainty on metallicity versus spectral SNR\@.  The top panel shows
the differences in [Fe/H] measured from the two different slitmasks.
These measurements are different by up to 0.8~dex because most of them
(those observed on both slitmasks) are based on partial exposures with
low SNRs.  The bottom panel shows the estimated uncertainty on
metallicity versus SNR\@.  These estimates are based on the full
exposures (coadded spectra from both slitmasks).  Therefore, the
bottom panel shows larger SNRs than the top panel.  The trend of error
with SNR is weaker than for radial velocities
(Figure~\ref{fig:vdiffsn}) because $\sigma_{\rm MC}$ for radial
velocities was calculated with Monte Carlo resampling of the observed
spectra.  Monte Carlo resampling was computationally prohibitive for
estimating uncertainties on [Fe/H]\@.  As a result, we relied in part
on $\sigma_{\rm covar}$ instead of $\sigma_{\rm MC,[Fe/H]}$ for
estimating the uncertainty on [Fe/H]\@.  This quantity comes from the
covariance matrix, which tends to underestimate uncertainty compared
to Monte Carlo techniques.  Therefore, $\sigma_{\rm sys,[Fe/H]}$
accounts for the majority of the uncertainty in most stars, and the
resulting error distribution shows only a weak trend with SNR\@.

In the following discussion, we have assumed that the
solar iron abundance is $12 + \log n({\rm Fe})/n({\rm H}) = 7.52$
\citep{sne92}.  Also, we discarded metallicity measurements with
uncertainties larger than 0.5~dex.

\begin{figure}[t!]
\centering
\includegraphics[width=\columnwidth]{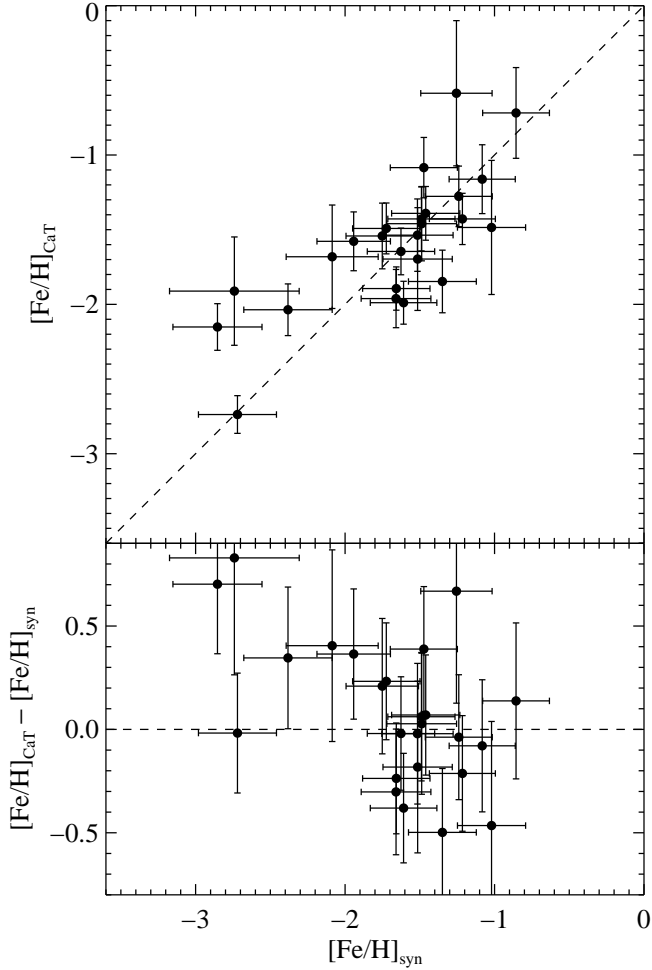}
\caption{Metallicities derived from the CaT versus metallicities
  derived from spectral synthesis.\label{fig:fehcat}}
\end{figure}

Another popular method of measuring metallicities of red giants from
low- to medium-resolution spectra is the empirical relation between
[Fe/H] and the EW of the CaT\@.  In order to compare CaT metallicities
to our synthetic spectral metallicities, we measured the EWs of the
CaT lines at 8542~\AA\ and 8662~\AA\@. We fitted Lorentzian functions
to each of the CaT lines.  We estimated uncertainties on EWs by Monte
Carlo resampling.  We added Gaussian random noise to each spectrum in
proportion to its error spectrum.  The amount of noise added to each
pixel was sampled from a Gaussian random distribution with a standard
deviation equal to that pixel's estimated Poisson noise.  We took the
error on EW as the standard deviation of 100 such realizations.  Then,
we applied \citeauthor{sta10}'s (\citeyear{sta10}) calibration between
CaT EW, absolute $V$ magnitude, and [Fe/H]\@.  This calibration was
based on a combination of Gaussian fits and numerical integration in
order to include the pressure-broadened wings of the CaT\@.  As a
rough substitute, we used the Lorentzian, which is a good
approximation to both the cores and the wings of the CaT lines.  We
transformed extinction-corrected $g_0$ and $r_0$ magnitudes to Cousins
$V_0$ magnitudes using \citeauthor{jor06}'s (\citeyear{jor06})
equations.  We computed errors on ${\rm [Fe/H]}_{\rm CaT}$ by
propagating the errors on EW and $V$.  We did not include the
uncertainty in the metallicity calibration because it is significantly
less than the substantial errors in EW\@.  Figure~\ref{fig:fehcat}
shows the comparison between [Fe/H] measured from the CaT and from
spectral synthesis for the stars where it was possible to measure
Lorentzian fits confidently.  Although the measurement uncertainties
are large, the agreement is good.  This comparison reinforces
credibility in our metallicity measurements based on \ion{Fe}{1}
lines.  For the rest of this discussion, we used synthetic
metallicities instead of the CaT metallicities because the synthetic
metallicities are more precise for two reasons.  First, the synthetic
metallicities do not require a conversion between photometric systems.
Second, \ion{Fe}{1} lines in the DEIMOS spectrum of a typical
moderately metal-poor red giant have more signal (sum of EWs) than the
combined CaT\@.

\begin{figure}[t!]
\centering
\includegraphics[width=\columnwidth]{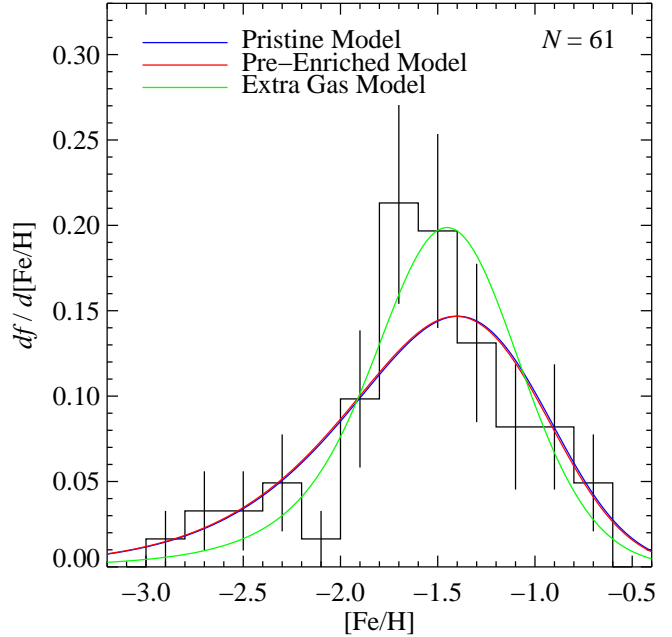}
\caption{Metallicity distribution for member stars.  Error bars are
  calculated with Poisson statistics.  Best-fitting, analytic models
  of chemical evolution are shown as colored curves.  The red and blue
  lines overlap almost exactly.\label{fig:fehhist}}
\end{figure}

Figure~\ref{fig:fehhist} shows the metallicity distribution of member
stars.  The metallicities range from $\rm{[Fe/H]} = \fehmin$ to
$\fehmax$.  The shape of the distribution (a peak with a longer
metal-poor tail than a metal-rich one) is typical of dwarf galaxies of
VV124's stellar mass \citep{kir11a}.  The bottom of
Table~\ref{tab:parameters} lists some additional shape characteristics
of the metallicity distribution.

We measured a mean metallicity of $\langle {\rm [Fe/H]} \rangle =
\fehmean \pm \fehmeanerr$.  This number agrees with previous
photometric estimates: ${\rm [Fe/H]} = -1.37$ \citep{kop08,tik10},
${\rm [Fe/H]} = -1.79$ \citep{jac11}, ${\rm [M/H]} = -1.5$ (in the
outskirts, \citeauthor*{bel11a}), and ${\rm [M/H]} = -1.4$ \citep[in
  the region $0.5' < R_{\epsilon} < 1.0'$, where $R_{\epsilon}$ is the
  elliptical radius,][]{bel11b}.  The metallicities measured by
\citet{jac11} and \citet{bel11a,bel11b} are consistent because
\citeauthor{jac11}\ measured [M/H] whereas
\citeauthor{bel11b}\ measured [Fe/H]\@.  \citet{jac11} and
\citet{bel11a,bel11b} also noted that the great majority of old RGB
stars have ${\rm [Fe/H]} \le -1$.
%Moreover, photometric metallicities can be biased by the presence of
%an age spread in the old stellar population that dominates the galaxy
%\citep{bel11b}.

%the latter two photometric estimates are themselves discrepant by
%0.29~dex.  Furthermore, \citeauthor*{bel11a} based their estimate on
%the outermost stars, 2.5' to 5.0' (1 to 2~kpc) from the center of the
%galaxy.  The radial metallicity gradient in VV124 (see below) ensures
%that \citeauthor*{bel11a}'s estimate of [Fe/H] for the entire galaxy
%would be slightly closer to our measurement.

\citet{kir11a} fitted three analytic chemical evolution models to the
metallicity distributions of eight dSph satellites of the Milky Way.
The Simple Model is also known as a Closed Box \citep{tal71}.  Its
single parameter is the effective yield, $p$.  Low effective yield
indicates that the galaxy lost metals.  Hence, we prefer the name
``Simple Model'' to ``Closed Box.''  The Pre-Enriched Model
\citep{pag97} is the Simple Model with the additional parameter
$\rm{[Fe/H]}_0$, which is the initial metallicity of the gas.  The
Extra Gas Model \citep{lyn75} allows for gas to fall into the galaxy
during star formation.  The infall rate is specifically chosen to
allow for an analytic solution to the metallicity distribution.  The
parameter $M$ quantifies the amount of infalling gas.  The Extra Gas
Model reduces to the Simple Model in the limit $M=1$.

We fitted the same models with the same maximum likelihood estimation
procedure as \citet{kir11a}.  The colored curves in
Figure~\ref{fig:fehhist} show the results.  The curves for the Simple
and Pre-Enriched Models nearly overlap because the Pre-Enriched Model
reduces to the Simple Model in the limit that $\rm{[Fe/H]}_0 =
-\infty$.  The maximum likelihood estimation could not constrain a
lower bound to $\rm{[Fe/H]}_0$.  Hence, the metallicity distribution
is consistent with initially metal-free gas.  The metallicity
distribution is more peaked than the Simple Model.  The Extra Gas
Model roughly accounts for this peak.  In fact, the Extra Gas Model is
$\probratio$~times more likely to describe the metallicity
distribution than the Simple Model.  Metallicity distributions
consistent with gas inflow are typical for dSphs with the stellar mass
of VV124 \citep{kir11a}.

The low metallicity and highly sub-solar effective yields ($p \approx
0.04~Z_{\sun}$) require that VV124 lost some of its metals
\citep[see][]{kir11b}, possibly in the form of supernova winds.
\citeauthor*{bel11a}\ also noted the role of gas flows into and out of
VV124\@.  The \ion{H}{1} gas distribution is asymmetric and suggestive
of gas flowing to or from the southeast of the galaxy's star-forming
center.  The densest concentration of gas is also offset to the
southeast, further supporting the bulk motion of gas along that
direction.  While the present gas motions did not affect the past star
formation that made the bulk of the RGB, the radio observations do
support the role of gas flows in shaping the stellar population of
VV124\@.

\begin{figure}[t!]
\centering
\includegraphics[width=\columnwidth]{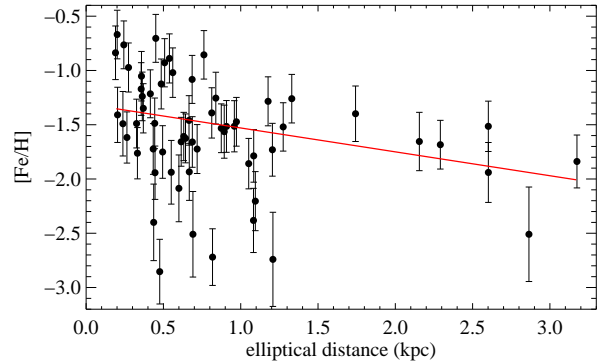}
\caption{Metallicity as a function of elliptical distance from the
  center of VV124\@.  The elliptical distance is the semi-major axis
  of the ellipse ($\epsilon = 0.44$) on which the star lies.  The red
  line is a least-squares fit.\label{fig:fehdist}}
\end{figure}

Radial metallicity gradients are typical of many galaxies, and many
dwarf galaxies are no exception
\citep[e.g.,][]{tol04,bat06,koc06,spo09,kol09a,kol09b,kir11a}.  If
gradients exist, they are always negative in the sense that
metallicity decreases with increasing radius.  Photometry strongly
indicated that the stars near the center of VV124 are younger and more
metal-rich \citep{tik10,jac11,bel11b}.  We confirmed a gradient of
$d{\rm [Fe/H]}/dr = \fehslope \pm \fehslopeerr~{\rm dex~kpc}^{-1}$.
Figure~\ref{fig:fehdist} shows the radial distribution of [Fe/H] along
with an illustration of the gradient.  Radial gradients can arise by
several mechanisms.  The most plausible explanation is that
metallicity increased with time as supernovae enriched the
interstellar medium with iron.  Star formation also became more
centrally concentrated because gas loss (due to supernova winds, for
example) preferentially expelled gas from the outer reaches of the
galaxy.  Therefore, the radial gradient arose because successive
generations of stars became simultaneously more metal-rich and more
centrally concentrated.

%%%%%%%%%%%%%%%%%%%%%%%%%%%%%%%%%
%%%%%%%%%   SECTION 7   %%%%%%%%%
%%%%%%%%%%%%%%%%%%%%%%%%%%%%%%%%%

\section{Discussion and Conclusions}

\begin{figure*}[t!]
\centering
\includegraphics[width=\textwidth]{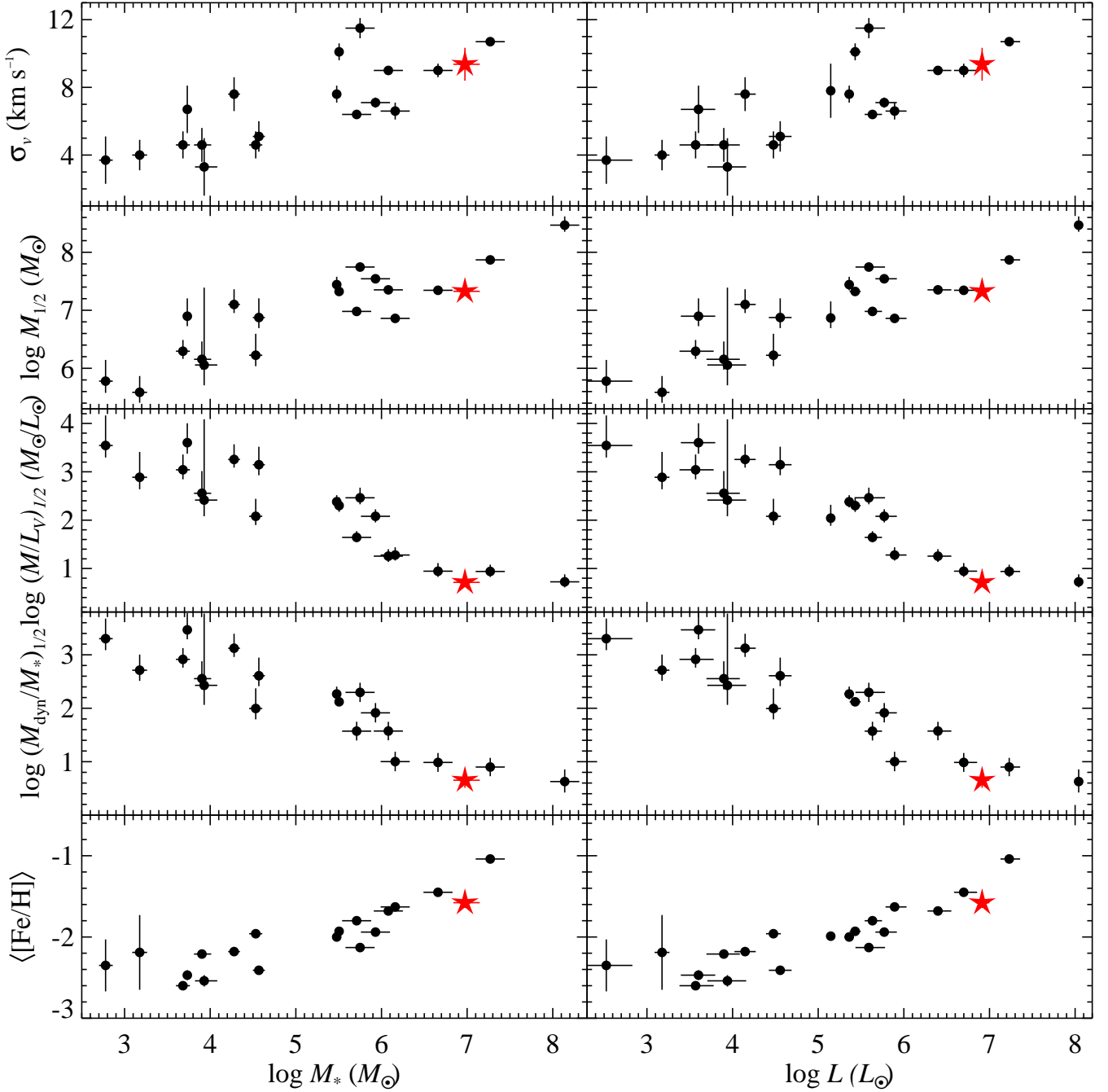}
\caption{Velocity dispersion, mass within the half-light radius,
  mass-to-light ratio within the half-light radius,
  dynamical-to-stellar mass ratio within the half-light radius, and
  average metallicity versus stellar mass (left) and luminosity
  (right) for dSph and dwarf elliptical galaxies in the Local Group.
  VV124 is represented by a red star.  Dynamical quantities
  ($\sigma_v$, $M_{1/2}$, $(M/L_V)_{1/2}$, and $(M_{\rm
    dyn}/M_*)_{1/2}$) were taken from \citet{wol10} and references
  therein for all galaxies except Bo{\"o}tes~I\@.  For Bo{\"o}tes~I,
  we adopted $\sigma_v$ from \citet{kop11}, and we adjusted its mass
  accordingly.  The stellar masses were taken from \citet{woo08} for
  the larger dSphs and \citet[][using the values derived with the
    \citeauthor{kro93}\ \citeyear{kro93} initial mass function]{mar08}
  for the ultra-faint dSphs.  The metallicities came from
  \citet{kir11a} except for Carina \citep{hel06} and Bo{\"o}tes~I
  \citep{mar07}.\label{fig:trends}}
\end{figure*}

Since its recent relocation to the edge of the Local Group, the dwarf
galaxy VV124 has received considerable scrutiny.  However,
measurements of its radial velocity have been inconsistent.  The
galaxy also seemed to possess a disk of stars.  We observed VV124 with
Keck/DEIMOS, and we determined that its radial velocity agrees with
\ion{H}{1} measurements.  The stellar velocities we measured show no
evidence of a disk rotating any faster than
$\avgvdifferr$~km~s$^{-1}$, but larger samples will be necessary to
place stricter limits on rotation.

We gave an estimate of the dynamical mass of VV124 and determined that
the mass of its stars alone does not account for the velocity
dispersion.  Therefore, VV124 contains dark matter.  Its mass-to-light
ratio agrees with dSphs of similar stellar mass.  The metallicity
distribution is also similar to dSphs of similar stellar mass.  The
average metallicity, metallicity dispersion, and shape of the
distribution are nearly indistinguishable from dSphs like Leo~I or
Fornax.  As is typical for large dSphs, the metallicity distribution
fits a chemical evolution model with infalling gas better than a
Closed Box Model.

The parameters we derived from new spectroscopy suggest that VV124 is
a dIrr far along on its transformation into a dSph.
Figure~\ref{fig:trends} shows that VV124 lies along the same trends
defined by other dSphs and dwarf elliptical galaxies in the Local
Group.  Thus, the galaxy is an excellent test case for models that
posit that dIrrs evolve into dSphs by the removal of gas and
randomization of stellar velocities.  This process may happen in
isolation or in the vicinity of a much larger galaxy, such as the
Milky Way.

The transformation of a dIrr into a dSph might proceed without the
benefit of a dense environment.  For example, VV124 in its present
form could have resulted from a merger of two dwarf galaxies $\sim
1$~Gyr ago \citep{don09,kaz11b}.  The merger would have disrupted a
disk and possibly triggered the starburst that \citet{jac11} observed.
Recent simulations showed that ultraviolet background radiation and
stellar feedback are sufficient to expel all of the gas from a dwarf
galaxy \citep{saw10}.  The challenge in these models is the formation
of disk-less dSphs.  It is possible that VV124 could have formed as a
dispersion-supported galaxy, but this scenario runs contrary to the
view that dwarf galaxies formed as disks from which supernova feedback
removed gas with low angular momentum \citep{gov10}.  Another
challenge is to explain the low gas fraction of VV124 relative to
other isolated dwarf galaxies.  In a sample of 39 dwarf galaxies with
stellar masses from \citet{woo08} and gas masses from \citet{gre03},
the only galaxies with lower gas fractions than VV124 ($f_{\rm gas} =
0.09$) are satellites of the Milky Way or M31, not isolated dwarfs
like VV124\@.

The tidal stirring model \citep{may01a,may01b} explains the
transformation from dIrrs to dSphs in the context of galaxy
environment.  This model posits that disky dIrrs lose their gas
through ram pressure stripping in a group environment.  The stars then
lose their rotational structure by tidal interactions.  Numerous
simulations address the transformation of dIrrs to dSphs by tidal
stirring \citep[e.g.,][]{kli09a,lok10,lok11,kaz11a}.  The stated goal
of most of these simulations is to construct a spheroid from a disk.
As a result, few studies address the kind of morphology peculiar to
VV124 (a spheroid with a very faint disk-like feature).  Nonetheless,
the intermediate stages of some tidal stirring simulations resemble a
spheroid with a disk.  The simulated tidally stirred dwarf galaxy in
Figure~12 of \citet{may01b} is a decent representation of VV124, but
the ``disk'' is actually tidal tails.  Our radial velocity
observations of stars in the wings of VV124 do not support their
interpretation as a disk.  Hence, they may be tidal tails.  However,
tidal tails near apocenter are expected to be elongated toward the
perturber \citep{kli09b}.  The disk- or tail-like feature observed in
VV124 must be nearly perpendicular to the Milky Way, and the great
circle that connects VV124 and M31 intersects the major axis of VV124
at an angle of $54\arcdeg$.  Therefore, the elongated photometric
features are unlikely to be tidal tails resulting from an interaction
with either of the large spiral galaxies of the Local Group.

Furthermore, VV124 is in a remote corner of the Local Group, highly
isolated from other galaxies.  \citet{kar09} concluded that VV124
resides outside the zero-velocity surface of the Local Group.  Because
its velocity relative to the Local Group's barycenter is positive, it
will never enter the group in the future.  Additionally,
\citeauthor*{bel11a}\ calculated that the intragroup medium around
VV124 is probably not dense enough to remove gas by ram pressure
stripping.  As a result, tidal stirring would require that VV124
passed near another galaxy and subsequently traveled $\sim 1$~Mpc to
its current position.  Traveling such a large distance after a tidal
interaction is not out of the question.  A dwarf galaxy that passed
within half of the virial radius of a massive halo could have traveled
up to several virial radii beyond that massive halo
\citep{lin03,lud09,tey12}.  As many as 10\% of all dark matter halos
are actually ejected subhalos that can be found as far away as four
virial radii from their former hosts \citep{wan09}.  VV124 qualifies
as a candidate subhalo ejected from either the Milky Way or M31.  Its
low velocity relative to the Local Group would require VV124 to be
near its apocenter.  Such a location would not be surprising given its
large distance from the nearest massive halo.  It also makes sense
that we would observe an ejected subhalo near its apocenter because
subhalos linger near their apocenters.  As a potentially ejected
satellite galaxy, VV124 could be an example of tidal stirring.  In
this scenario, a brief, high-speed passage near the Milky Way or M31
removed most of its gas, and the violent change in tidal force near
pericenter disrupted its stellar disk.  A single passage is sufficient
for ram pressure stripping to remove most or all of the gas in a
galaxy of VV124's circular velocity \citep{may06}, but most
simulations of tidal stirring require multiple pericentric passages to
transform a disk into a spheroid \citep{kaz11a}.

The tidal stirring explanation requires special circumstances.  First,
VV124 had to have been ejected from the Milky Way or M31, possibly in
a multi-body encounter.  Second, the stellar disk would have to be
disrupted in just one pericentric passage, or VV124 would have had to
acquire a large amount of orbital energy in a multi-body encounter on
a subsequent pericentric passage.  Third, the tidal tails---if indeed
the photometric wings are tidal tails---would have to somehow point in
a direction away from the former host galaxy.

While our observations support the interpretation of VV124 as a galaxy
of a type between dIrr and dSph, they do not explain how it came to be
that way.  We look forward to future simulations that will manufacture
a digital replica of VV124: an isolated, mostly spheroidal dwarf
galaxy that somehow lost most of its gas but retained wispy, minimally
rotating wings.

%Regardless, \citeauthor{kop08}'s (\citeyear{kop08}) re-discovery of
%VV124 was very recent.  The coming years will certainly witness much
%more detailed study of this enigmatic galaxy.  Whether it truly
%possesses a disk and whether it is a dIrr in the process of becoming a
%dSph can be resolved with additional deep spectroscopy of its young
%and old, stellar and gaseous components.

\acknowledgments We are grateful to the many people who have worked to
make the Keck Telescope and its instruments a reality and to operate
and maintain the Keck Observatory.  The authors wish to extend special
thanks to those of Hawaiian ancestry on whose sacred mountain we are
privileged to be guests.  Without their generous hospitality, none of
the observations presented herein would have been possible.

We thank Marla Geha for providing radial velocity template spectra
observed with DEIMOS and for her generous assistance in measuring
radial velocities.  We also thank the referee for a detailed report
that greatly improved this manuscript.  Support for this work was
provided by NASA through Hubble Fellowship grant 51256.01 awarded to
ENK by the Space Telescope Science Institute, which is operated by the
Association of Universities for Research in Astronomy, Inc., for NASA,
under contract NAS 5-26555.  JGC thanks NSF grant AST-0908139 for
partial support.

{\it Facility:} \facility{Keck:II (DEIMOS)}

\clearpage
\renewcommand{\thetable}{\arabic{table}}
\setcounter{table}{1}
\begin{turnpage}

\begin{deluxetable}{lccccccccccc}
\tablewidth{0pt}
\tablecolumns{12}
\tablecaption{Target List\label{tab:catalog}}
\tablehead{\colhead{ID} & \colhead{RA (J2000)} & \colhead{Dec (J2000)} & \colhead{$g$} & \colhead{$r$} & \colhead{Masks\tablenotemark{a}} & \colhead{SNR (\AA$^{-1}$)} & \colhead{$v_r$ (km~s$^{-1}$)} & \colhead{EW(\ion{Na}{1}~8190) (\AA)} & \colhead{[Fe/H]} & \colhead{Member?} & \colhead{Reason\tablenotemark{b}}}
\startdata
13585   &   09 15 09.6 & +52 49 52.0 & $24.351 \pm 0.030$ & $23.295 \pm 0.043$ & 1 &   4.1 &      \nodata      &      \nodata     &     \nodata      & N &               G  \\
16765   &   09 15 10.8 & +52 48 34.1 & $25.600 \pm 0.038$ & $23.667 \pm 0.046$ & 1 &   4.9 &      \nodata      &      \nodata     &     \nodata      & N &              Bad \\
11688   &   09 15 11.8 & +52 50 12.4 & $25.268 \pm 0.029$ & $23.365 \pm 0.025$ & 1 &  15.5 &      \nodata      &      \nodata     &     \nodata      & N &               G  \\
16355   &   09 15 11.8 & +52 48 50.4 & $24.183 \pm 0.011$ & $23.125 \pm 0.016$ & 1 &   4.1 &      \nodata      &      \nodata     &     \nodata      & N &              Bad \\
16895   &   09 15 13.4 & +52 48 29.4 & $24.198 \pm 0.021$ & $23.337 \pm 0.032$ & 1 &   2.2 &      \nodata      &      \nodata     &     \nodata      & N &              Bad \\
3705    &   09 15 15.4 & +52 51 59.3 & $24.651 \pm 0.015$ & $23.217 \pm 0.018$ & 1 &  25.6 & $ +31.5 \pm  2.7$ &  $3.21 \pm 0.24$ &     \nodata      & N &        $v_r$ Na  \\
16598   &   09 15 16.9 & +52 48 40.7 & $23.009 \pm 0.006$ & $21.553 \pm 0.007$ & 1 &  50.9 & $ -68.1 \pm  2.4$ &  $2.22 \pm 0.06$ &     \nodata      & N & $v_r$ Na Bright  \\
16798   &   09 15 17.5 & +52 48 32.7 & $22.361 \pm 0.006$ & $20.841 \pm 0.007$ & 1 &  68.8 & $ -25.5 \pm  2.3$ &  $2.45 \pm 0.04$ &     \nodata      & N &       Na Bright  \\
7035    &   09 15 17.9 & +52 50 58.8 & $22.720 \pm 0.010$ & $22.720 \pm 0.013$ & 2 &  16.5 &      \nodata      &      \nodata     &     \nodata      & N &               G  \\
7917    &   09 15 19.8 & +52 50 49.5 & $24.287 \pm 0.012$ & $23.183 \pm 0.017$ & 2 &  13.0 & $ -40.1 \pm  4.8$ &      \nodata     & $-1.51 \pm 0.23$ & Y &                  \\
\nodata & \nodata & \nodata & \nodata & \nodata & \nodata & \nodata & \nodata & \nodata & \nodata & \nodata & \nodata \\
\enddata
\tablerefs{Identifications, coordinates, and photometry from \protect \citeauthor*{bel11a}.}
\tablenotetext{a}{Number of DEIMOS masks on which the object was observed.}
\tablenotetext{b}{Reasons for non-membership.  $v_r$: Inappropriate radial velocity.  Na: Spectrum shows strong \ion{Na}{1}~$\lambda$8190 doublet.  G: Spectrum shows emission lines or redshifted Ca H and K lines, indicating that the object is a galaxy.  Bright: Target is brighter than the TRGB\@.  Bad: Spectral quality was insufficient for radial velocity measurement.}
\tablecomments{(This table is available in its entirety in a machine-readable form in the online journal. A portion is shown here for guidance regarding its form and content.)}
\end{deluxetable}

\clearpage
\end{turnpage}

\end{document}